\documentclass[aps,pra,reprint,superscriptaddress,longbibliography]{revtex4-2}

\usepackage{amsmath, amsthm, amsfonts, amssymb,amstext}
\usepackage{mathptmx}
\usepackage{bm}
\usepackage{xfrac}
\usepackage{mathrsfs}
\usepackage{latexsym}
\usepackage{array}

\usepackage{soul}

\usepackage{graphicx}

\usepackage{amsmath,amssymb,amsfonts}

\usepackage{enumitem}

\usepackage{xcolor}

\usepackage{soul,xcolor}

\usepackage{tikz,tkz-euclide}

\usepackage[colorlinks=true,citecolor=blue,linkcolor=blue,urlcolor=blue]{hyperref}

\newcommand{\abs}[1]{\left|#1\right|}

\let\ni\noindent
\usepackage{physics}
\usepackage{verbatim}
\usepackage{array}
\usepackage{tabularx}

\usepackage{tikz,ifthen} 
\usetikzlibrary{shapes,arrows,positioning,automata,backgrounds,calc,er,patterns,matrix}

\begin{document}
\begin{abstract}
We study the finite-density phases of a $\mathbb{Z}_2$ lattice gauge theory (LGT) of interconnected loops and dynamical $\mathbb{Z}_2$  charges. The gauge-invariant Wilson  terms, accounting for the  magnetic flux threading each loop,  correspond to simple  two-body Ising interactions in this setting. Such terms control the interference of charges tunneling around the loops, leading to dynamical Aharonov-Bohm (AB) cages that are delimited by loops threaded by a $\pi$-flux. The latter can be understood as $\mathbb{Z}_2$ vortices, the analog of visons in two dimensional LGTs, which become mobile by adding quantum fluctuations through an external electric field. In contrast to a semi-classical  regime of static and homogeneous AB cages, the mobile visons  can self-assemble leading to AB cages of different lengths  depending on the density of $\mathbb{Z}_2$ charges and the interplay of magnetic and electric terms. Inside these cages,  the individual charges get confined into tightly-bound charge-neutral pairs, the  $\mathbb{Z}_2$ analogue of mesons. Depending on the region of parameter space, these tightly-bound mesons can propagate  within dilute AB-dimers that virtually expand and contract, or else move by virtually stretching and compressing an electric field string. Both limits lead to a Luttinger liquid described by a constrained integrable model. This phase is separated from an incompressible Mott insulator where mesons belong to  closely-packed AB-trimers. In light of recent trapped-ion experiments for a single $\mathbb{Z}_2$ loop, these phases could be explored in  future experiments.
\end{abstract}

\title{Dynamical Aharonov-Bohm cages and tight meson confinement in a $\mathbb{Z}_2$-loop gauge theory}

\author{Enrico C. Domanti}

\email{Enrico.Domanti@gmail.com}
\affiliation{Quantum Research Centre, Technology Innovation Institute, Abu Dhabi, UAE}

\affiliation{Dipartimento di Fisica e Astronomia, Via S. Sofia 64, 95127 Catania, Italy}

\affiliation{INFN-Sezione di Catania, Via S. Sofia 64, 95127 Catania, Italy}

\author {Alejandro Bermudez}

\affiliation{Instituto de Física Teorica, UAM-CSIC, Universidad Autonoma de Madrid, Cantoblanco, 28049 Madrid, Spain}

\author {Luigi Amico}

\affiliation{Quantum Research Centre, Technology Innovation Institute, Abu Dhabi, UAE}

\affiliation{Dipartimento di Fisica e Astronomia, Via S. Sofia 64, 95127 Catania, Italy}

\affiliation{INFN-Sezione di Catania, Via S. Sofia 64, 95127 Catania, Italy}

\maketitle

\date{\today}

\begin{section}{Introduction}
Gauge theories are ubiquitous in many areas of physics, ranging from the realm of fundamental interactions \cite{Peskin1995intro,yang1954conservation}, to the low-energy physics of condensed matter systems \cite{wen2017zoo,Sachdev2016emergent}. Dynamical gauge fields mediate the interactions of quantum matter and are responsible for remarkable non-perturbative phenomena, being confinement a prominent example~\cite{greensite2020introduction}. The impossibility of observing isolated particles
with a net gauge charge according to the underlying local symmetry group \cite{Peskin1995intro}, first realised in asymptotically-free non-Abelian gauge theories \cite{gross1973ultraviolet}, can actually also appear in simpler Abelian models, such as quantum electrodynamics in $D=1+1$ dimensions~\cite{schwinger1962gauge}. While the dimensional reduction allows for a more detailed quantitative understanding of this phenomenon~\cite{COLEMAN1975267,coleman1976more}, a full non-perturbative analysis of finite-density regimes and long real-time propagation lies beyond our current  capabilities. The discretization of  gauge theories on a space-time lattice does provide a well-defined route to address several non-perturbative problems in the field, such as the phase structure of quantum chromodynamics. However, our most sophisticated  numerical analysis of lattice gauge theories (LGTs) are limited by the so-called sign problem~\cite{nagata2022finite} and the harsh entanglement scaling properties~\cite{RevModPhys.80.517,RevModPhys.93.045003}, such that a plethora of quantitative questions remain unanswered.

\begin{figure}
    \centering
    \includegraphics[width=0.8\linewidth]{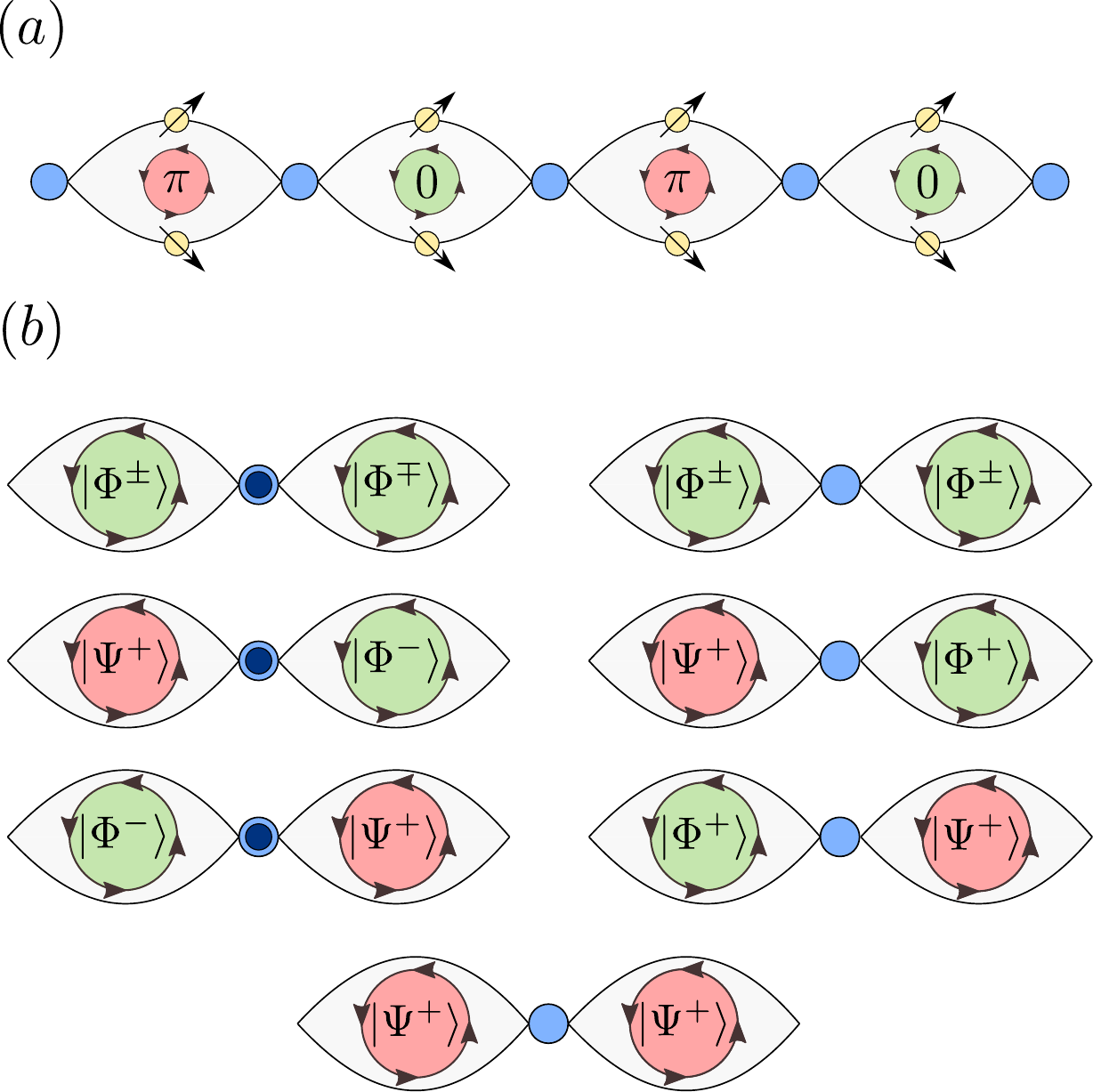}
    \caption{\textit{$\mathbb{Z}_2$ Loop-chain}. (a) Hardcore bosons live in the sites of a one-dimensional lattice. Two links depart from each site in a loop geometry and host spin-$\frac{1}{2}$ gauge-field variables $\sigma^\alpha_1$ and $\sigma^\alpha_2$. Each loop encloses a gauge flux $\sigma^x_1 \sigma^x_2$ that can take values in $\{0,\pi\}$. The hopping amplitude between sites connected by a loop in a $\pi$-flux state vanishes due to destructive Aharonov-Bohm interference. (b) Gauge invariant configurations in the spin-$1$ reduction of the loop-chain model, where $|\Phi^\pm\rangle$ and $|\Psi^+\rangle$ are common eigenstates of the spin-$1$ operators $(S^x)^2$ and $(S^z)^2$. Occupied sites are filled in dark-blue.}
    \label{fig:loop_chain}
\end{figure}

With the advent of quantum technologies, an alternative strategy has been identified~\cite{Feynman1982}, which searches for scalable protocols to synthesize LGTs on either digital or analogue quantum simulators \cite{buchler2005atomic,Weimer2010rydberg,zohar2011confinement,zohar2012simulating,zohar2013cold,zohar2013simulating,Zohar2016quantum,banerjee2012atomic,banerjee2013atomic,banuls2020review,Banuls2020simulating,tagliacozzo2013optical,Tagliacozzo2013simulation,Wiese2013ultracold,dalmonte2016lattice,Aidelsburger2021cold,domanti2024floquet,Bazvan2024synthetic}. Such an endeavour has led to the experimental observation of gauge-invariant dynamics, confinement and string-breaking phenomena~\cite{Martinez2016real,Dai2017four,klco2018quantum,Schweizer2019floquet,Kokail2019self,surace2020lattice,klco2020su2,mil2020scalable,Yang2020observation,zhao2022thermalization,Atas2021su2,bauer2021simulating,nguyen2022digital,ciavarella2021trailhead,mildenberger2022probingconfinementmathbbz2lattice,rahman2021su2,ciavarella2022preparation,wang2022observation,atas2023simulating,farrell2023preparations,farrell2023preparations2,su2023observation,charles2024simulating,zhang2023observation, Cochran:2024rwe,de2024observationstringbreakingdynamicsquantum,gonzalezcuadra2024observationstringbreaking2,liu2024stringbreakingmechanismlattice,oxford_z2_gauge}. Despite being still far from a fully-fledged  quantum simulation of the Standard Model of particle physics, which will likely require a large-scale fault-tolerant quantum computer, these simpler  quantum simulators can already provide useful insights, especially when addressing real-time  and finite-density phenomena. LGTs built on $\mathbb{Z}_2$ gauge groups constitute a prototypical example of such models, which are of great interest in various contexts, ranging from the understanding of confinement \cite{borla2020confined,borla2021gauging,kebric2021confinement,gonzalez2020robust,gao2022fractional,Gazit2017emergent,fradkin1979phase}, to topological phases of matter and spin liquids \cite{wen2013topological,wen2007quantum,tupitsyn2010topological,kitaev2003fault}, opening the way to new collective effects\cite{gonzalez2018strongly,gonzalez2019symmetry,Gonzalez-Cuadra2019}.

Let us consider a $D=(1+1)$ dimensional $\mathbb{Z}_2$ LGT in which the dynamical charges correspond to hardcore bosons that can be created or annihilated by $a^\dagger_{i},a^{\phantom{\dagger}}_{i}$ on the lattice sites $x_{i}=id$, where $d$ is the lattice spacing and $i\in\mathbb{Z}_L$ with $L$ being the number of sites. The boson operators are constrained by $a_i^{\phantom{\dagger}\!\! 2}=0=a_i^{{\dagger} 2}$, such that there cannot be double occupancies, and their tunneling will be  mediated by $\mathbb{Z}_2$ gauge fields encoded by Ising spins $\sigma^\alpha_{i_\ell}$ on the  links $x_{i_\ell}=i_\ell d$ with $i_\ell=i+{1}/{2}$, where $\alpha\in\{x,y,z\}$ specifies the particular Pauli operator. The $\mathbb{Z}_2$ LGT on a chain has been thoroughly studied previously~\cite{borla2020confined,kebric2021confinement,kebric2024confinement,kebric2024mean}. In the presence of a non-zero electric field strength $h$, which leads to a  potential  between a pair of background test charges  growing linearly with their distance $V(r)\sim V_0 r$, this model exhibits confinement also when the charges become dynamical \cite{borla2020confined,kebric2021confinement}. Such property is reflected in the exponential decay of the gauge-invariant one-body Green's function, indirectly showing that there are no charged quasi-particles in the low-energy spectrum. Confinement in the $\mathbb{Z}_2$ chain has been shown to survive for any particle density~\cite{kebric2024mean}, and even at non-zero temperatures~\cite{kebric2024confinement}. For very large electric fields $h$, the particles get tightly bound into charge-neutral pairs  at neighboring sites  connected by a short electric-field string. The emergent dynamics of these bound pairs is governed by an integrable model that, in the continuous limit, has a Luttinger liquid (LL) behavior with a power-law decay for the corresponding Green's function~\cite{borla2020confined}. Hence, in contrast to the charged particles, these charge-neutral $\mathbb{Z}_2$ ``mesons'' propagate as  boson fields with a certain anomalous dimension, governing the physics of the low-energy  spectrum of the theory. As one lowers the electric field, even if  the confinement is not so tight and the exact mapping to the integrable model is no longer valid, the $\mathbb{Z}_2$ chain remains in a confined LL phase, albeit with an anomalous dimension and power-law decay that changes.

In our work, we explore a minimal extension of this model that brings in a neat connection to quantum Hall physics, where the interplay of dynamical matter and {\it static} gauge fields has been the object of intense investigations. In this case, the gauge fields provide a background under which the charged particles propagate. A notable effect in such systems is that charged particles in 2d and quasi-1d lattices can localize even in the absence of disorder, as a result of Aharonov-Bohm interference~\cite{PhysRev.115.485}. This phenomenon is known as Aharonov-Bohm (AB) caging~\cite{vidal1998aharonov}, as particles are locked inside restricted regions of space. This leads to localised eigenstates and a flat-band spectrum for specific values of the background magnetic flux that pierces each elementary plaquette. The competition of Aharonov-Bohm caging and inter-particle interactions has been widely addressed in the literature: while interactions are disruptive for the stability of cages, they can be responsible for a rich physical phenomenology \cite{vidal2000interaction,PhysRevB.82.184502,doucot2002pairing,PhysRevLett.107.150501,PhysRevX.7.031057,cartwright2018rhombi,danieli2020many,roy2020interplay}.

With our work, we introduce the notion of {\it dynamical AB caging},  in which gauge fields are no longer static classical configurations, but rather dynamical quantum variables.  Indeed, we shall see that a fundamental interplay between  AB caging and  confinement physics arises. Specifically, also motivated by recent experimental progress in the implementation of a $\mathbb{Z}_2$ loop with trapped-ion devices~\cite{oxford_z2_gauge}, we introduce a minimal LGT generalising the above $\mathbb{Z}_2$ chain, and grounding for the phenomenon of the aforementioned dynamical AB caging through gauge fields. For this purpose, we add an extra Ising spin on each link of the system we sketched above, such that the charges have now two tunneling paths along the bonds $b\in\{1,2\}$, each mediated by the gauge-field parallel transporter  $\sigma^z_{b,i_\ell}$. This setup can then be naturally represented as a linear chain of loops, with two $\mathbb{Z}_2$ gauge fields living on either arm  of each loop - see Fig.~\ref{fig:loop_chain}. These minimal plaquettes thus enclose a dynamical $\mathbb{Z}_2$-valued flux, whose magnetic energetic contribution stems from a simplified Wilson plaquette term that only requires a two-body Ising interaction to attain gauge invariance. As we shall see, such a $\mathbb{Z}_2$-flux, taking values $0$ or $\pi$, can indeed set the sought dynamical AB caging. {In the context of fractionalization and  $\mathbb{Z}_2$ LGTs with dynamical matter in  $D=2+1$ dimensions, $\pi$ fluxes are referred to as visons for vortex Ising excitations ~\cite{PhysRevB.62.7850,Sachdev_2019}. They appear as particle-like vortex excitations localised to a single plaquette of the two dimensional lattice and can only be excited by a non-local string that connects the plaquette to one of the boundaries. By analogy, we shall also refer to these localised $\pi$ fluxes as visons for our $\mathbb{Z}_2$-loop chain.} While here visons cannot condense to yield confinement-deconfinement transitions, we will demonstrate that phase transitions of a different nature can occur, as arising from the interplay of the particle filling, visons proliferation and AB cages. 

This article is organized as follows. In sections \ref{sub:model} and \ref{sub:sp1} the Hamiltonian of the model is introduced and its symmetries are thoroughly discussed. In Sec.\ref{sub:dmrg} a novel approach to numerically solve the system with DMRG, while restricting to subsectors of the Hilbert space in which all the symmetries of the model are fixed, is presented and relies on a specific mapping. The main results of this work are then presented in sections \ref{sub:h_0}, \ref{sub:small_h} and \ref{sub:mott}, where both numerical and analytical results are discussed to corroborate our analysis of the system. We summarize our results and draw our conclusions in Sec.\ref{sec:conclusions}.
\end{section}

\begin{section}{Methods}
\begin{subsection}{The model}
\label{sub:model}
The  Hamiltonian of the $\mathbb{Z}_2$-loop chain, for $\hbar=1$, reads
\begin{equation}
\label{eq:hamiltonian}
    \begin{aligned}
     H &= \frac{t}{2}\sum_{i,b}  \big(a_i^\dagger \sigma^z_{b,i_\ell}a^{\phantom{\dagger}}_{i+1} + {\rm H.c.}\big) \,  + \frac{h}{2}\sum_{i,b} \, \sigma^x_{b,i_\ell} 
         + \frac{J}{2} \, \sum_i \sigma^z_{1,i_\ell} \sigma^z_{2,i_\ell},
    \end{aligned}
\end{equation}
where we have introduced the tunneling strength $t$, and the electric  $h$ and magnetic $J$ coupling.
The tunneling of hardcore bosons is assisted by a Pauli operator along $z$ that acts as a parallel transporter, enforcing the $\mathbb{Z}_2$ gauge invariance of the matter dynamics. Pauli operators along $x$, instead, play the role of a $\mathbb{Z}_2$ electric field $E^{\phantom{x}}_{b,i_\ell} = (1 +\sigma^x_{b,i_\ell})/2$. In the electric-field basis, the states $\ket{\pm_{b,i_\ell}} =(\ket{\uparrow_{b,i_\ell}} \pm \ket{\downarrow_{b,i_\ell}})/{\sqrt{2}}$ stand for the presence/absence of  an electric-field line connecting  two neighboring matter sites. The Ising coupling $W_{\bigcirc_{i_\ell}} =  \sigma^z_{1,i_\ell}\sigma^z_{2,i_\ell}$ can be interpreted as a Wilson plaquette term that quantifies the 'magnetic flux'  piercing the loop that connects sites $i$ and $(i+1)$. Indeed, for a vanishing electric field, the total phase acquired by a boson circulating the loop can be expressed as an effective flux ${\rm exp}\{\rm i {\Phi}_{\rm B}\}= \langle W_{\bigcirc_{i_\ell}}\rangle$. As a result, ferromagnetic loop orderings are $0$-flux configurations, while anti-ferromagnetic ones yield a  $\pi$-flux. Since $\pi$-flux states correspond to vanishing eigenvalues of $S^z_{i_\ell}= \frac{1}{2} (\sigma^z_{i_\ell,1}+\sigma^z_{i_\ell,2})$, such gauge-field loop configurations lead to a perfect destructive AB interference, inhibiting the boson  tunneling  to a neighboring loop. In contrast with the classical AB cages, we note that the interplay of the  interference and particle dynamics will be affected by quantum fluctuations and by the specific  filling.

The Hamiltonian \eqref{eq:hamiltonian} has a global $U(1)$ symmetry associated to the conservation of the total particle number, which we control through a chemical potential $\mu$ via   $H\to H-\mu \sum_i n_i$ with $n_i=a_i^\dagger a_i^{\phantom{\dagger}}$.
The $\mathbb{Z}_2$ gauge invariance of the Hamiltonian~\eqref{eq:hamiltonian} results from  $\comm{H}{G_i} = 0 \, \forall i$, in which $G_i = \prod_b \sigma^{x}_{b,i_\ell-1} (-1)^{n_i} \prod_b \sigma^{x}_{b,i_\ell}$ are the local symmetry generators. When  considering an open loop chain terminating with a pair of matter sites,  the local generators of the gauge group at its ends should read $G_1 = (-1)^{n_1} \prod_b \sigma^{x}_{b,1_\ell}$ and $G_L = \prod_b \sigma^{x}_{b,(L-1)_\ell} (-1)^{n_L}$.
\end{subsection}

\begin{subsection}{Additional local symmetry and reduction to spin-$1$}
\label{sub:sp1}
Besides its $\mathbb{Z}_2$ gauge symmetry, the Hamiltonian \eqref{eq:hamiltonian} is also  invariant under the local exchange of the two Ising operators of each loop: $\sigma^\alpha_{1,i_\ell} \leftrightarrow \sigma^\alpha_{2,i_\ell}$. Let us define the total spin in a loop as $\boldsymbol{S}_{i_\ell} = ( \boldsymbol{\sigma}_{1,i_\ell} + \boldsymbol{\sigma}_{2,i_\ell})/2$, so that the $\mathbb{Z}_2$-loop Hilbert space can be seen as the direct sum of a spin-$0$ (singlet) and a spin-$1$ (triplet) representation of SU(2). There is thus an additional local conservation of the Casimir operator $\boldsymbol{S}^2_{i_\ell}$, which has eigenvalues $s_{i_\ell} (s_{i_\ell} + 1)$, with $s_{i_\ell} \in \{0,1\}$. The total Hilbert space thus decouples into sectors corresponding to each possible set $\{\dots, s_{i_\ell-1}, s_{i_\ell},s_{i_\ell+1}, \dots\}$. Whenever a loop is in a singlet $\pi$-flux configuration, which corresponds to $s_{i_\ell} =0$ and to a loop Bell state $|\Psi^-_{i_\ell}\rangle=(\ket{\uparrow_{1,i_\ell}\downarrow_{2,i_\ell}}-\ket{\downarrow_{1,i_\ell}\uparrow_{2,i_\ell}})/\sqrt{2}$, the chain gets effectively broken, as the loop resides in a so-called dark state that decouples from the dynamics. Moreover, the associated $\pi$-flux inhibits any possible tunneling connecting the two partitions. Therefore, an arbitrary distribution of singlets leads to the chain being fragmented into a set of independent sub-chains, which are instead composed only of triplet  bonds that allow for non-trivial dynamics. 

Without loss of generality,  we can thus restrict our analysis to one of these $s_{i_\ell} = 1$ sub-chains. When expressed in terms of the total spin, the Hamiltonian reads
\begin{equation}
    \label{eq:spin1red}
     H = t\sum_{i}  \big(a_i^\dagger S^z_{i_\ell}a^{\phantom{\dagger}}_{i+1} + {\rm H.c.}\big) \,  + {h}\sum_{i} \, S^x_{i_\ell} 
         + {J}  \sum_i \big(S^{z}_{i_\ell}\big)^{\!2},
\end{equation}
where we have neglected an irrelevant
constant term. The model in Eq.~~\eqref{eq:spin1red} now describes  a LGT on a chain, but with gauge degrees of freedom being  spin-$1$ operators. Let us remark that this model differs from the so-called quantum link models~\cite{chandrasekharan1997quantum}, which can also include a higher-spin representations of the gauge fields to preserve a local U(1) gauge symmetry. In our case, the symmetry is still the discrete $\mathbb{Z}_2$ group, having effective local generators that now read 
\begin{equation}
\label{eq:gauss_generator}
G_i = P^x_{i_\ell-1}  (-1)^{n_i} P^x_{i_\ell},\hspace{2ex} P^x_{i_\ell} = 2(S_{i_\ell}^x)^2-1
\end{equation}
together with the corresponding deformation at the edges for open boundary conditions. We will work in the neutral gauge sector, such that the physical space $\ket{\psi}\in\mathcal{H}_{\rm phys}\subset\mathcal{H}$ is stabilised by the Gauss' law operators $G_i\ket{\psi} = +\ket{\psi} \, \forall i$. In this case, considering a finite chain ending with sites, one can see that the total parity of hardcore bosons is actually fixed to be even, as $P_{\text{hb}} =  (-1)^{\sum_i n_i} = \prod_i G_i = 1$. We note that the loop chain model \eqref{eq:hamiltonian} is the simplest extension of the $\mathbb{Z}_2$ chain in which the gauge-field dynamics is not entirely fixed by the matter fields. Indeed, the non linearity of the Gauss' law in terms of the gauge field operators, which is quadratic in the spin-1 reduction of the Hamiltonian, Eq. \eqref{eq:spin1red}, makes it impossible to integrate them out in favour of a pure matter model, in contrast to the $\mathbb{Z}_2$ chain \cite{borla2020confined}. 

In the  spin-$1$ language, the Wilson plaquette for each loop becomes $W_{\bigcirc_{i_\ell}}= 2(S^z_{i_\ell})^2 -1$, and thus has a $\mathbb{Z}_2$-valued spectrum $\pm 1$
corresponding to the $0$-flux ($+1$) and $\pi$-flux ($-1$) configurations. Equivalently, the total electric field in each loop is controlled by the $x$ component of the total spin  $E_{i_\ell} =\sum_bE^{\phantom{x}}_{b,i_\ell}=  1 + S^x_{i_\ell}$. From here onward, we will adopt the notation $\displaystyle{\ket{0_ 
\Uparrow} = \ket{\uparrow \uparrow},\ket{\pi} = (\ket{\uparrow \downarrow} + \ket{\downarrow \uparrow})/{\sqrt{2}},\ket{0_\Downarrow} = \ket{\downarrow \downarrow}}$ for the $m=1,0,-1$ eigenstates of $S^z$ that label  $0$-flux and   a $\pi$-flux states respectively. For convenience, we will also use the Bell-pair notation $\ket{\Phi^{\pm}} = (\ket{0_\Uparrow} \pm \ket{0_\Downarrow})/{\sqrt{2}}$, $\ket{\Psi^+}=\ket{\pi}$,  recalling that the missing Bell pair $\ket{\Psi^-}=(\ket{\uparrow \downarrow} - \ket{\downarrow \uparrow})/{\sqrt{2}}$ is the dark state that completely decoupled from the dynamics. We note that the role of gauge symmetry and AB interference in these entangled flux states has been realised in recent trapped-ion experiments~\cite{oxford_z2_gauge}. The Bell states are common eigenstates of $(S_{i_\ell}^x)^2$ and $(S^z_{i_\ell})^2$, with $(S^x_{i_\ell})^2 |\Phi^+_{i_\ell}\rangle = |\Phi^+_{i_\ell}\rangle $, $(S^x_{i_\ell})^2 |\Psi^+_{i_\ell}\rangle  = |\Psi^+_{i_\ell}\rangle$, $(S^x_{i_\ell})^2 |\Phi^-_{i_\ell}\rangle  = 0$ and $(S_{i_\ell}^z)^2 |\Phi^\pm_{i_\ell}\rangle = |\Phi^\pm_{i_\ell}\rangle$, $(S^z)^2 |\Psi^\pm_{i_\ell}\rangle = 0$. We also point out that the state $|\Phi^-_{i_\ell}\rangle$ is a $0$-flux state which also has a well defined value of the total electric field $E_{i_\ell} |\Phi^-_{i_\ell}\rangle= |\Phi^-_{i_\ell}\rangle$.  The gauge-invariant configurations which are allowed by the Gauss law are depicted in Fig.~\ref{fig:loop_chain}.
\end{subsection}
\begin{subsection}{DMRG encoding}
\label{sub:dmrg}
\begin{figure}
    \includegraphics[width =0.85 \linewidth]{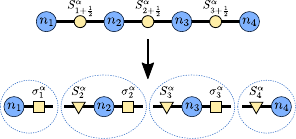}
    \caption{\textit{Schematics of the encoding for DMRG}. The bonds of the spin-$1$ LGT in Eq.~\eqref{eq:spin1red} are broken in half. Super-sites (dashed circles) enclose the original matter sites of the chain and two "half-links", except at the edges, where one "half-link" is dropped as the original chain terminates with sites. In each super-site, the left "half-links" host spin-$\frac{1}{2}$ degrees of freedom, while spin-$1$ operators live in those on the right.}
    \label{fig:encoding}
\end{figure}

The interest in low-dimensional LGTs with simplified Abelian groups lies in the expectations that they could  be amenable to quantum simulation in the near term. See~\cite{Martinez2016real,Dai2017four,klco2018quantum,Schweizer2019floquet,Kokail2019self,surace2020lattice,klco2020su2,mil2020scalable,Yang2020observation,zhao2022thermalization,Atas2021su2,bauer2021simulating,nguyen2022digital,ciavarella2021trailhead,rahman2021su2,ciavarella2022preparation,wang2022observation,atas2023simulating,farrell2023preparations,farrell2023preparations2,su2023observation,charles2024simulating,zhang2023observation, Cochran:2024rwe,de2024observationstringbreakingdynamicsquantum,gonzalezcuadra2024observationstringbreaking2,liu2024stringbreakingmechanismlattice,oxford_z2_gauge} for recent experimental progress.  Even if phenomena related to long-time dynamics lies beyond numerical capabilities, the  existence of an entanglement area law~\cite{Hastings2007an} for the reduced dimensionality suggests that one may use numerical simulations based on matrix product states (MPS)~\cite{verstraete2008matrix} to explore the full finite-density phase diagram of the model. This will shed light on the role of the dynamical AB cages and the gauge-invariant couplings, guiding our understanding of finite-density effects reported in the following sections. Before delving into the details, let us present our approach to deal with the local $\mathbb{Z}_2$ symmetry using MPS.

The numerical results of this paper are mainly obtained using DMRG algorithms \cite{schollwock2011the,fishman2022the,itensor-r0.3} in subspaces of fixed gauge sector $G_i=1$ and particle density $\nu = N/L$. A direct use of DMRG for Eq.~~\eqref{eq:spin1red} requires that an MPS will carry independent physical indices ('sites') for both hardcore bosons and spin-$1$ gauge-fields. To each matter site we can associate quantum numbers corresponding to the local particle density $n_i$. When quantum numbers can be associated to each physical index of an MPS, it is possible to resort to quantum number conserving algorithms that enforce the corresponding symmetry, such that, in our case, we can impose particle-number conservation within the DMRG iterations. On the other hand, the extensively many constraints imposed by the gauge generators~\eqref{eq:gauss_generator}, each involve operators over three neighboring sites of an MPS, one for matter and two for gauge-fields. 
One may consider adding Hamiltonian penalty terms such as $H\to H-\lambda \sum_i G_i$, which suppress unwanted contributions from other gauge-sectors in the low-energy regime, but these are hard to control \cite{Felser2020two}. Another possible strategy would be to integrate out the matter degrees of freedom
by making use of Gauss' law $G_i\ket{\psi} = +\ket{\psi}$ $\forall i$, with the generators in Eq.~~\eqref{eq:gauss_generator}. In the physical subspace, this would lead to a spin-$1$ model in a similar spirit to~\cite{borla2020confined,borla2021gauging}, with non-local terms that follow from the mapping of the particle density  onto  domain-wall operators $n_i \to ( 1 - P^x_{i_\ell-1} P^x_{i_\ell})/2$, which count the number of magnetic kinks connecting two neighboring ferromagnetic domains
-- see Supplemental Material. In this case, a fine-tuning of a chemical potential term $-\mu( 1 - P^x_{i_\ell-1} P^x_{i_\ell})/2$ would be needed to fix the total number of particles in the system.

In order to enforce both  Gauss' law  and particle number conservation without resorting to penalty or chemical potential terms, we rewrite the  model \eqref{eq:spin1red} as follows. First, every bond operator is split in two $S_{i_\ell}^\alpha\mapsto\sigma^\alpha_i,S^\alpha_{i+1}$. Then, 'super-sites' are introduced that enclose the original matter sites of the chain and the neighboring 'half-links'. The half-links on the right of each site host Ising spin degrees of freedom, while spin-$1$ operators live on the left -- see Fig.~\ref{fig:encoding}. The Hamiltonian is then rewritten as $H=\mathcal{P} \tilde{H} \mathcal{P}$, where
\begin{equation}
\label{eq:dmrg_mapped}
    \tilde{H} = t\sum_i  (a_i^\dagger \sigma_i^z S_{i+1}^z a^{\phantom{\dagger}}_{i+1} + {\rm H.c.})  + h\sum_i S_i^x + J \,\sum_i (S_i^z)^2   
\end{equation}
\ni still displays a local $\mathbb{Z}_2$ gauge symmetry with generators that are local on the super-sites $\tilde{G}_i = P^x_i (-1)^{n_i} \sigma_i^x$, $\tilde{G}_1 = (-1)^{n_1} \sigma_1^x$, $\tilde{G}_L = P^x_L (-1)^{n_L}$. 
\begin{figure}
    \centering
    \includegraphics[width=0.9\linewidth]{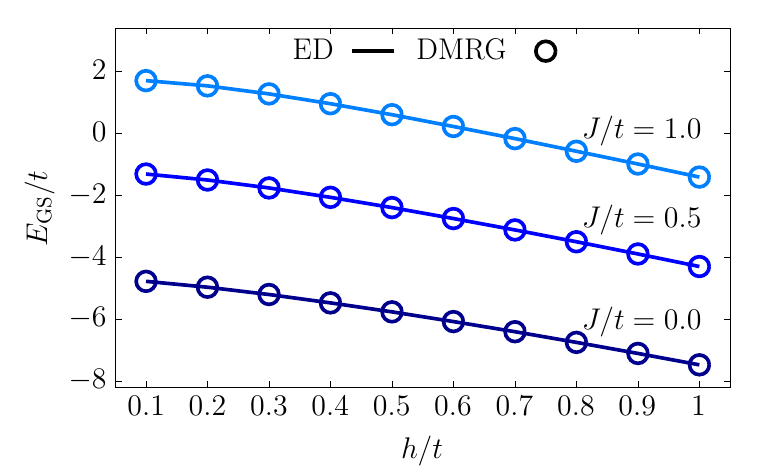}
    \caption{\textit{ED-DMRG comparison}. The results obtained by using DMRG with the devised encoding are compared to ED applied on Eq.~\eqref{eq:spin1red}, for a chain of $L=9$ sites and $N=6$ particles. The groundstate energies obtained with the two methods with varying $h/t$ are plot for $J/t = 0.0,0.5,1.0$.}
    \label{fig:comparison}
\end{figure}
An MPS will now have physical indices corresponding to each super-site, whose local Hilbert space we restrict to the sub-sector in which $\tilde{G}_i = 1$, by removing all states that do not satisfy these constraints. Moreover, the particle density $n_i$ also has support on each super-site, so that no chemical potential term is needed to fix the total number of particles. The projectors $\mathcal{P} = \prod_i (1+\sigma_i^x P^x_{i+1})/2$ impose the condition that
\begin{equation}
\label{eq:dmrg_constraint}
    \sigma_i^x = P^x_{i+1} \, ,
\end{equation}

\ni ensuring that the dynamics of $\tilde{H}$ in the sub-space constrained by $\tilde{G}_i = 1$ is reduced to that of Eq.~\eqref{eq:spin1red} in the neutral gauge sector. Indeed, $\sigma^x_i$ is introduced to act as a tracker of $P^x_{i+1}$ on the next super-site. When the constraints in Eq.~~\eqref{eq:dmrg_constraint} are fulfilled, the condition $\tilde{G}_i = 1$ will return the original Gauss' law $G_i = 1$ -- see Eq.~~\eqref{eq:gauss_generator}. The action of $S^z_{i+1}$, resulting in $P^x_{i+1} \to - P^x_{i+1}$, then requires $\sigma^x_i$ to change simultaneously -- see Eq.~~\eqref{eq:dmrg_mapped}. We remark that the projected Hamiltonian will still consist only of local terms and nearest-neighbour couplings between super-sites. Moreover, only states satisfying Eq.\eqref{eq:dmrg_constraint} will give a non-zero contribution to the energy. Nevertheless, a generic MPS will still have projections over the unwanted states and, since the constraints have support on neighboring super-sites, an energetic penalty would be required to ensure that such states do not appear in the ground-state after DMRG. As a solution, we construct a global symmetry operator $O = \sum_i O_i$, whose density $O_i$ only involves single-site operators and is such that a specific symmetry sector does not contain any of the unwanted states. Taking $O_i = 2^{i-2} P_i^x - 2^{i-1} \sigma_i^x$, we can write
\begin{equation}
\label{eq:dmrg_global}
    O = \sum_{i=1}^{L-1} O_i = \sum_{i=1}^{L-1} 2^{i-1} (P_{i+1}^x - \sigma_i^x) \, .
\end{equation}

 \ni The operator $O$ commutes with
 $\mathcal{P}^\dagger \tilde{H} \mathcal{P}$ by construction and $\gamma_i= P_{i+1}^x - \sigma_i^x$ has eigenvalues $\{-2,0,2\}$.  Restricting to the subspace in which $O=0$, all the aforementioned constraints  are automatically enforced, as none of the states violating the conditions \eqref{eq:dmrg_constraint} belongs to this symmetry sector -- see SM. As a figure of merit of the reliability of the above method, we show that Exact Diagonalization (ED) results based on Eq.~\eqref{eq:spin1red} and DMRG results based on Eq.~\eqref{eq:dmrg_mapped} coincide - see Fig.~\ref{fig:comparison}.
\end{subsection}
\end{section}

 \begin{figure}
     \centering
     \includegraphics[width= \linewidth]{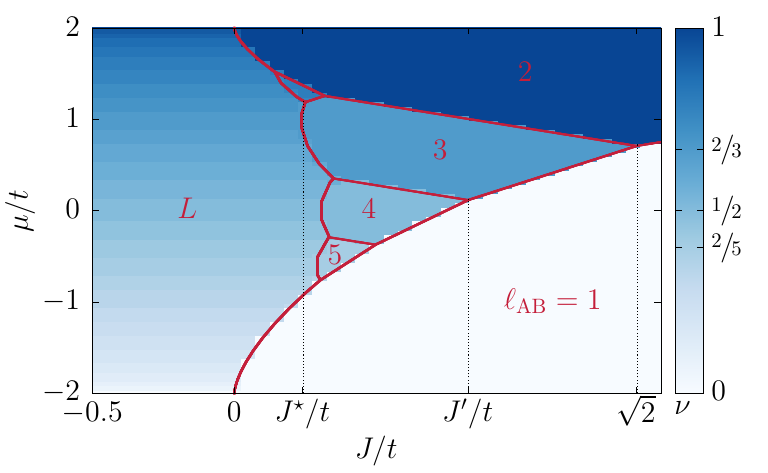}
     \caption{\textit{Phase-diagram at $h=0$}. Contour plot of the filling fraction $\nu=N/L$ as a function of $J/t$ and $\mu/t$, for a chain of $L=60$ sites. Red lines delimit areas of constant average gauge flux $\overline{W} = \sum_i \langle W_{\bigcirc_{i_\ell}}\rangle/(L-1)$ in which the chain is divided into clusters of size $\ell_{\rm AB}$ indicated by the red numbers inside each region. For instance, the  region on the left corresponds to the un-partitioned chain of $\ell_{\rm AB}=L$ connected sites and no visons, whereas that in the bottom right corner is broken at every site $\ell_{\rm AB}=1$ by completely filling the loops with visions. Except for the  small area with cages of size $\ell_{\rm AB}=5$, which  host four particles each,  the remaining arrangement of AB cages with $\ell_{\rm AB}\in\{1,2,3,4\}$ contain  two particles each. Notice that the reference values $J^\star/t = {3\sqrt{3}}/{\pi} - \sqrt{2} $, $J'/t = \sqrt{5}- \sqrt{2}$ and $J/t = \sqrt{2}$  can be calculated analytically in the thermodynamic limit- see Supplemental Material for details.}
     \label{fig:classical_unfixed}
 \end{figure}
 
\begin{section}{Results}
\begin{subsection}{Self assembly of Aharonov-Bohm cages at $h=0$}
\label{sub:h_0}
    For $h=0$, the electric field term drops out of the Hamiltonian \eqref{eq:spin1red} and the remaining hopping and flux terms locally commute. As a result, $(S^z_{i_\ell})^2$ is locally conserved, and the Hilbert space results to be fragmented into sectors labeled by the eigenvalues of $(S^z_{i_\ell})^2$ at each link of the chain. Following a similar logic to what has been discussed in the previous section, we conclude that the chain will get partitioned into all possible combinations of AB cages of different lengths, each surrounded by a pair of links in the $\ket{\pi}$ state, which supports a perfect AB destructive interference such that the particles that reside within the cage cannot escape. We will then be left with a set of independent AB cages, each hosting only $0$-flux bond states and a number of particles that will search for the minimum energy configuration consistent with the AB constrained length.
 \begin{figure}
     \centering
     \includegraphics[width=0.9\linewidth]{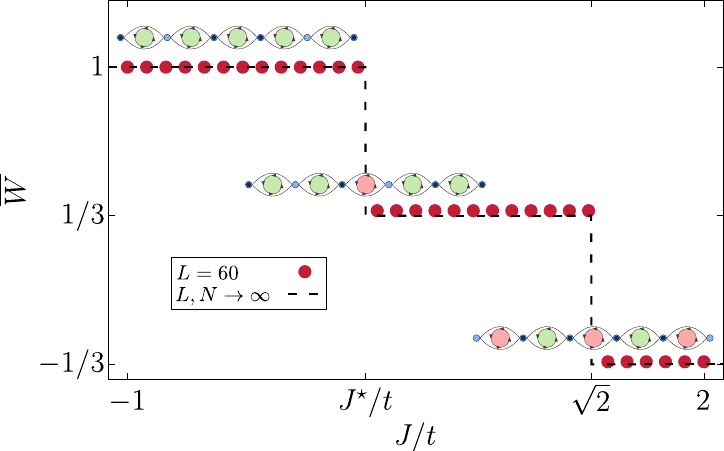}
     \caption{\textit{Chain-partitioning.} We plot the averaged expectation value of the magnetic Wilson  operators $\overline{W} = \sum_i \langle W_{\bigcirc_{i_\ell}}\rangle/(L-1)$, which provides a sense on how the number of vison changes with the parameters and, thus, on the formation of AB cages at $h=0$ and filling $\nu=\frac{2}{3}$. The fragmented chains contain AB cages separated from one another by $\pi$-flux loops (colored in red).  By increasing $J/t$, we observe the transition  from a fully-connected chain to an AB-trimer covering and, finally, to  an AB-dimer configuration. In the thermodynamic limit, the transition points are given by $J^\star/t = {3 \sqrt{3}}/{\pi} - \sqrt{2}$ and $J/t =  \sqrt{2}$.}
     \label{fig:partitions}
 \end{figure}
 Every AB cage will only be allowed to host either zero or an even number of particles, as we recall that there is a  parity constraint dictated by  Gauss' law. Finding the ground state of the system reduces to a classical problem, which consists of comparing the energies of all the possible different partitionings of the full chain by allowing for all possible distributions of $\pi$ fluxes, and accounting for all the allowed particle distributions that individually minimize the kinetic energy in each AB cage - see Supplemental Material.
 
 The behaviour of the system with changing $J/t$ and $\mu/t$ is presented in Fig.~\ref{fig:classical_unfixed}. The boundaries in red enclose regions of a constant average gauge flux $\overline{W} = \sum_i \langle W_{\bigcirc_{i_\ell}}\rangle/(L-1)$, which provides indirect information on the nature of the chain partitioning. We find that the lattice does not break into partitions for $J<0$, as the magnetic Wilson term favours in this case a ferromagnetic ordering of the spins, leading to an overall 0-flux. For positive values, we also find a vanishing flux  below a  pronged curve that separates the phase with a fully-connected chain from one with AB cages and clustering of particles. Below the transition line, the number of particles decreases in steps of two with decreasing $\mu$. When crossing the curve, we observe the self-assembling of caged configurations favoured energetically: due to destructive AB interference, particles are locked in clusters covering the entire chain according to a periodic crystalline structure. Five main commensurate coverings of the lattice emerge, with cages of sizes $\ell_{\rm AB} = 1,2,3,4,5$. Partitions of length $\ell_{\rm AB}=1$ correspond to independent empty sites and to a total of zero particles in the lattice, appearing in the bottom right corner of the figure. For all the other values of $\ell_{\rm AB}$, except for a small area that corresponds to partitions of $\ell_{\rm AB} = 5$ sites that contain four particles, each $\ell_{\rm AB}$-sized cage is filled with just two particles. As a result, the filling fraction $\nu$ is fixed to the value $\nu = {2}/{\ell_{\rm AB}}$.
 
 To the best of our knowledge, and in contrast to the previous results on AB caging in  static and homogeneous flux brackgrounds \cite{vidal1998aharonov,vidal2000interaction,doucot2002pairing}, this is the first study reporting a spontaneous fractionalization of the system by the  nucleation  of $\pi$ fluxes, sometimes referred to as visons in the context of particle-like deconfined excitations of $\mathbb{Z}_2$ LGTs~\cite{PhysRevB.62.7850,Sachdev_2019}. Here, these visons get arranged forming ordered patterns with a periodicity that depends on the  filling and the competition of the various microscopic terms. We note that, for a chain of finite size $L$, depending on the number of sites, coverings with a single kind of clusters may become incommensurate with the size of the lattice, resulting in mixed configurations. Such effect is expected to disappear in the thermodynamic limit $L\to\infty$.

Let  us now fix the total number of particles, and explore this phenomenon in more detail for the specific filling fraction $\nu = {2}/{3}$. Resorting to numerical methods, we find the following behaviour - see Fig.~\ref{fig:partitions}: for any $J<J^\star$, where in the thermodynamic limit $J^\star/t = {3\sqrt{3}}/{\pi} - \sqrt{2}$, the system attains its minimum energy by remaining fully connected without AB cages or visons. Strictly at vanishing electric field  $h=0$, our hardcore bosons hopping on a whole chain then correspond to deconfined excitations in a metallic phase. Increasing the magnetic coupling to  $J^\star < J < \sqrt{2} t$, pairs of particles get caged in three-sites  cages (AB trimers) that completely cover the chain. Finally, for even larger couplings  $J \geq \sqrt{2}t$, we find that the system lowers its energy by creating smaller cages of two sites (AB dimers), in which two particles are arranged into a tightly confined 'meson'. In this case, due to the specific particle number $N=2L/3$,  one cannot densely cover the lattice with these meson-filled AB dimers, and the groundstate shall intersperse AB dimers  with other empty sites/clusters. We note that this comes with a large degeneracy, corresponding to all the possible arrangements of the diluted AB dimers. 

We point out that different values of the filling fraction could set configurations other than the ones discussed above. For instance, for $\nu = {2}/{3}$, it is not commensurable with the total number of particles and sites in the chain to replace any number of trimers with  cages of four sites and two particles each. Such AB fourmers configurations can instead be found  at half-filling $\nu = {1}/{2}$ - see Supplemental Material (SM).
\end{subsection}
\begin{subsection}{Interference-assisted tight confinement of mesons}\label{sub:small_h}
Let us now switch on the electric field $h>0$,  introducing quantum fluctuations into the loop fluxes that compete with the AB caging, and tend to restore the tunneling among the disconnected partitions of the chain. In particular, visons start to move along the chain, which is accompanied by the rearrangement  of AB cages that carry with them the particles that were locked inside. Whilst the effect of $h>0$ tends to be disruptive for the AB caging, at small $h/t$ and $h/J$ configurations of constant average gauge flux $\overline{W}$ and filling fraction $\nu$ are resilient to quantum fluctuations and the filling fraction displays a staircase behaviour with changing $\mu/t$ - see Fig.\ref{fig:filling_cut}. Plateaux of fixed $\nu$ correspond to incompressible configurations, being characterized by $\displaystyle{\kappa = {\partial \langle N\rangle }/{\partial \mu} = 0}$, at which the average flux $\overline{W}$ is also constant, indirectly showing the persistence of caged configurations. The size of these plateaux tends to shrink as the strength of the fluctuations controlled by $h$ is increased.
\begin{figure}[h!]
    \centering
    \includegraphics[width=0.95\linewidth]{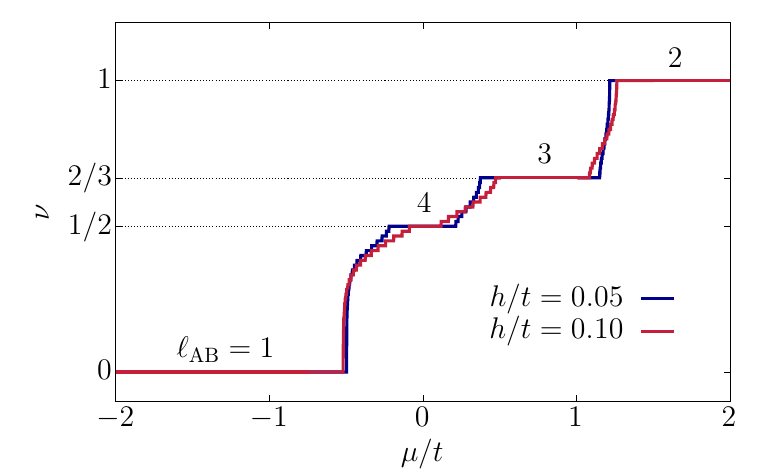}
    \caption{\textit{Staircase behaviour of the filling fraction}. We plot the filling fraction $\nu$ as a function of the chemical potential $\mu/t$ at small values of the electric field strength $h/t = 0.05, 0.10$ and at $J/t = 0.45$. The length of the chain is $L=120$. Plateaux at the fillings $\nu = 0,1/2,2/3,1$ denote incompressible configurations of the ground state associated to a constant value of the average flux $\overline{W}$ and thus to the presence of cages of average size $\ell_{\rm AB} \sim 1,4,3$ and $2$ respectively, as indicated in the labels above each plateaux.}
    \label{fig:filling_cut}
\end{figure}

    We stress that, at small values of $h/t$ and $h/J$, most of the incompressible configurations found in the previous semi-classical  ground states are still present - see Fig.~\ref{fig:mott_lobes}. In fact, we find that cages of size $\ell_{\rm AB}=1,2,3,4$ survive quantum fluctuations, while those of size $\ell_{\rm AB} =5$ become unstable. As in the  semi-classical case, we find that for ground states with single-site AB cages  $\ell_{\rm AB}=1$, the system has a vanishing particle filling $\nu=0$. Otherwise,  all the remaining AB cages host two particles each, such that  $\nu = {2}/{\ell_{\rm AB}}$. We find that all these phases are incompressible since $\kappa = 0$ - see Fig.~\ref{fig:mott_lobes}, and spontaneously break the lattice translational symmetry down to the subgroup of $\ell_{\rm AB}$-site translations. These type of transitions is reminiscent of the Peierls' instabilty in low-dimensional metals, which breaks translational invariance into the subgroup of $2$-site translations leading to an  insulator in a dimerised chain \cite{peierls1955quantum}. In our case, we find periodic arrangement of visons and, in the thermodynamic limit, the ground state is expected to be $\ell_{\rm AB}$-fold degenerate. This degeneracy  corresponds to all the possible non-equivalent coverings of the lattice, namely those that cannot be related by $\ell_{\rm AB}$-site translations. 
Being characterized by a staircase behavior of $\kappa$,  transitions between different incompressible phases are first-order. 
\begin{figure*}
    \centering
    \includegraphics[width=\linewidth]{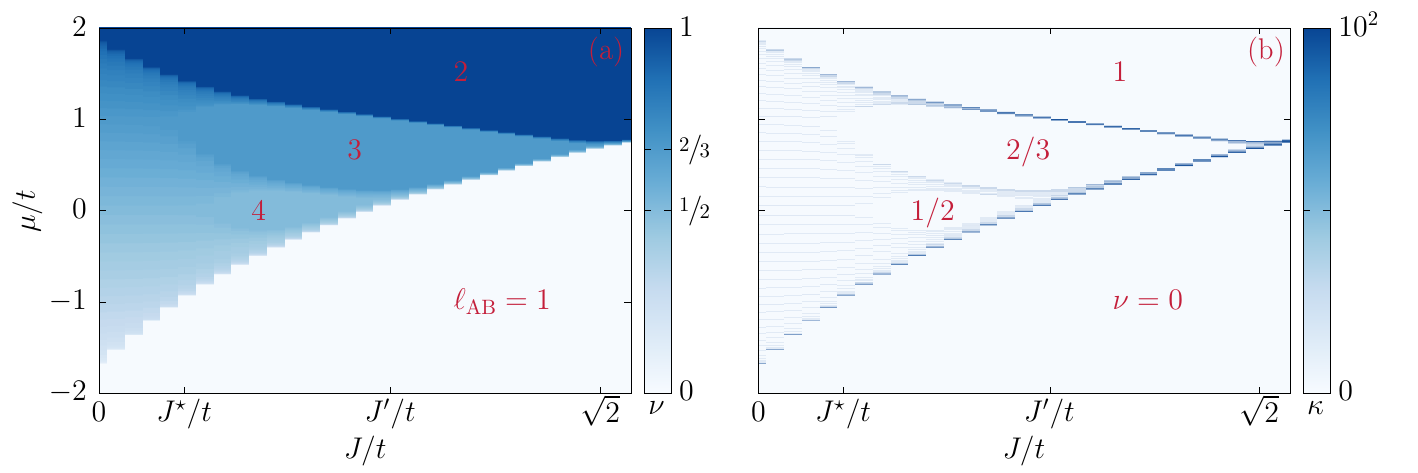}
    \caption{\textit{Incompressible $\mathbb{Z}_{\ell_{\rm AB}}$ phases}. In panel (a) we plot the filling fraction $\nu$ at $h/t=0.05$, for a chain of $L=120$ sites. At small values of the electric field, quantum fluctuations of the flux restore the hopping among separate partitions. Nevertheless, regions of fixed filling fraction $\nu$ and flux are stabilized at $\nu = 0$, $\frac{1}{2}$, $\frac{2}{3}$ and $1$, corresponding to incompressible phases in which the chain is fully partitioned into single empty sites, fourmers, trimers and dimers respectively, as indicated by the labels inside each area that correspond to the average size of the AB cages $\ell_{\rm AB}$. As a result, the lattice translational symmetry is broken to the $\mathbb{Z}_{\ell_{\rm AB}}$ subgroup. In panel (b) we show the compressibility $\kappa = \frac{\partial N}{\partial \mu} \Big\rvert_J$, demonstrating that Mott insulating regions are incompressible areas at fixed filling $\nu$, as indicated by the red labels in each lobe, and are separated by lines of diverging $\kappa$, at which the filling fraction changes abruptly.}
\label{fig:mott_lobes}
\end{figure*}

As noted above, in the limit of small $h/t$, $h/J$, quantum fluctuations can assist the movement of AB cages  and the particles thereby enclosed. This effect can be effectively described via a Schrieffer-Wolff perturbative transformation of Eq.~~\eqref{eq:spin1red}: we focus on the caged regimes in which the lattice is broken in clusters of $\ell_{\rm AB}$ sites containing two particles each, far from the values of $J/t$ at which cages of different lengths become degenerate. A non-zero 
electric field can induce virtual processes that expand and subsequently contract the AB cages by creating and destroying visons through $S_{i_\ell}^x |\Phi^+_{i_\ell}\rangle \to |\Psi^+_{i_\ell}\rangle$ and $S_{i_\ell}^x |\Psi^+_{i_\ell}\rangle \to |\Phi^+_{i_\ell}\rangle$. The successive expansion and contraction of an AB cage, which result from creating and destroying visons at its opposite edges, yield an effective hopping process. Conversely, this second-order process taking place at only one of its sides would simply lower the energy. These energy-lowering contributions are suppressed when a single $\pi$-flux loop is localised between neighbouring cages, which thus results in an effective repulsive interaction. In this case, second-order processes affecting this boundary vison cause the two $\ell_{\rm AB}$ cages to fuse into a single $2 \ell_{\rm AB}$ cage, and then break apart again. Taking into account both effective hoppings and interactions, we find that this dynamics is described by an extended Bose-Hubbard model, where an additional constraint   accounts for the spatial extent of each AB cage - see SM. The effective Hamiltonian reads
\begin{equation}
\label{eq:effham}
    H_{\ell_{\rm AB}} = \mathcal{P}_{\ell_{\rm AB}} \!\sum_i \left( - t^{\phantom{\dagger}}_{\ell_{\rm AB}} (b^\dagger_i b^{\phantom{\dagger}}_{i+1} + {\rm H.c.}) + V^{\phantom{\dagger}}_{\ell_{\rm AB}}  n_i^{\phantom{\dagger}} n^{\phantom{\dagger}}_{i+\ell_{\rm AB}} \right) \!\!\mathcal{P}_{\ell_{\rm AB}}.
\end{equation}
In this equation, we have introduced  bosonic operators $b^\dagger_i (b^{\phantom{\dagger}}_i)$, which create (annihilate) an entire $\ell_{\rm AB}$-sized AB cage that starts at site $i$. Hence, they involve the creation  (annihilation) of two particles in the lowest two-particle eigenstate available to a $\ell_{\rm AB}$-site sub-chain, where all the links enclosed by the cage are forced to $0$-flux ($\pi$-flux) states according to the Gauss' law -- see Fig.~\ref{fig:bosons}. Such operators are gauge-invariant meson creators(annihilators) as they can be written as linear combinations of two particle $a_i^{\phantom{\dagger}},a_j^{\phantom{\dagger}}$ ($a_i^\dagger,a_j^\dagger$) operators appropriately separated by strings of loop operators to fulfill the Gauss' law constraints. For instance, in the simplest case of AB dimers $\ell_{\rm AB}=2$ , we have $b_{i} = a_i \, |\Psi^+_{i_\ell}\rangle\langle \Phi^-_{i_\ell} |a_{i+1}$. Let us also note that the projectors $\mathcal{P}_{\ell_{\rm AB}}$ impose a cluster hardcore constraint $b^\dagger_i b^\dagger_{i+n} = 0$ that forbids the occurrence of pairs of doubly-occupied sites within any site in the cage $\forall n \in \{0,\ell_{\rm AB}-1\}$. Hence, only a single meson can reside in each AB cage.
\begin{figure}[h!]
    \centering
    \includegraphics[width=\linewidth]{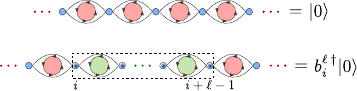}
    \caption{\textit{Schematic representation of the bosonic cages}. The state $\ket{0}$ is a fully polarized vacuum in which all the loops are in the $\ket{\pi}$ state and empty matter sites are disconnected from one another. The bosonic operator $b^{\ell \, \dagger}_i$ creates two particles, delocalized inside a cage between sites $i$ and $i+\ell-1$ (colored in shaded blue), in the lowest two-particle energy state for a subchain of length $\ell$. All link states are "flipped" to zero-flux configurations, in accordance to the Gauss law constraints.}
    \label{fig:bosons}
    \end{figure}
In the cases of relevance to our current discussion, namely those in which only two particles are hosted by each cage, the original density $\nu$ and the filling fraction of these mesons $\nu_{\rm m}(\ell_{\rm AB})$ are related via the expression $\displaystyle{\nu_{\rm m}(\ell_{\rm AB}) = {\nu}/{2}}$. Due to the constraints imposed by $\mathcal{P}_{\ell_{\rm AB}}$, the specific value $\bar{\nu}_{\rm m}(\ell_{\rm AB}) = {1}/(1+\ell_{\rm AB})$ is equivalent to  half filling, such that one cage exists on every other site. In a similar way, $\displaystyle{\tilde{\nu}_{\rm m}(\ell_{\rm AB}) = {1}/{\ell_{\rm AB}}}$ can be regarded as the fully-filled regime, since the projection constraints would not allow for additional clusters to be added. In this last case, the model \eqref{eq:effham} is forced into one of the configurations discussed above in which the chain is completely covered by the cages and the system is found in an incompressible phase, with the lattice translational invariance being broken to its $\mathbb{Z}_{\ell_{\rm AB}}$ subgroup- see Fig.~\ref{fig:mott_lobes}.

The model in Eq.~\eqref{eq:effham}  can be mapped to a  spin-${1}/{2}$ XXZ chain with a constraint that forbids spin configurations in which two spin-up states are less than $\ell_{\rm AB}$ sites apart. Such models have been exactly solved using Bethe-Ansatz  for any  value of $\ell_{\rm AB}$~\cite{alcaraz1999exactly}. The possible phases turned out to be  independent of $\ell_{\rm AB}$, leading to an equal-time Green's function that displays an algebraic decay $\langle b^ \dagger_i b_{i+r}\rangle\sim r^{-\beta}$, with  $\beta$ being dependent on the meson filling $\nu_{\rm m}(\ell_{\rm AB})$. At half filling for the  mesons, the effect of the above repulsive Hubbard interactions is enhanced, and one finds a quantum phase transition at ${\delta_{\rm c} = {V_{\ell_{\rm AB}}}/{2 \, t_{\ell_{\rm AB}}} = 1}$ from a gapless Luttinger Liquid phase ($\delta \leq 1$) to a gapped Mott-insulating phase ($\delta > 1$)~\cite{alcaraz1999exactly,borla2020confined}. 
\begin{figure}
    \centering
    \includegraphics[width=\linewidth]{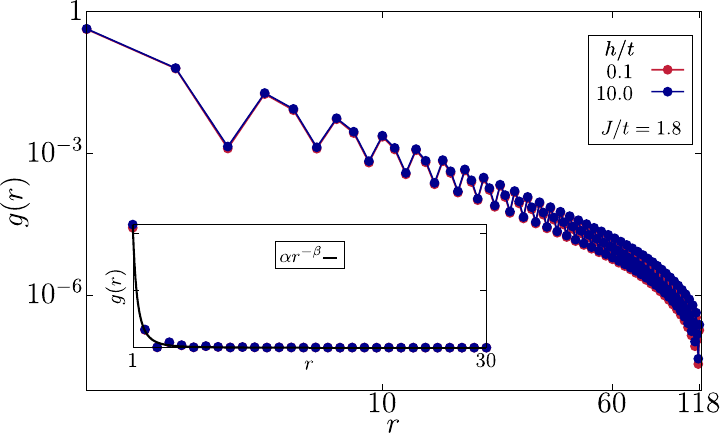}
    \caption{\textit{Dimer correlator}. The correlator $g(r) = \langle b_i^\dagger b_{i+r} \rangle$, with $b_i$ corresponding to the annihilator for a dimer, is shown in the opposite regimes of small ($0.1$) and large ($10.0$) values of $\frac{h}{t}$, at the filling $\nu = \frac{2}{3}$ and at the magnetic coupling $J/t = 1.8$. Given that the model \eqref{eq:spin1red} is defined on a chain of length $L=120$, dimer operators will be defined on a chain of length $L-1$. The main plot in log-log scale makes it evident that the algebraic decays in the two regimes are almost superimposed and confirms that the effective description provided by Eq.~\eqref{eq:effham} for $\ell_{\rm AB} = 2$ is valid in both cases. In the inset, the same plot is in linear scale up to $r=30$ for a better visualization. The data points are fitted to $g(r) = \alpha \, r^{-\beta}$. For $h/t = 0.1$($10.0$), we find $\alpha \sim 0.42$($0.43$) and $\beta \sim 2.87$($2.86$).}
    \label{fig:dimercorr}
\end{figure}

Below, we will discuss how the above exact results can be exploited to get insights about the phase diagram of Eq.~(\ref{eq:spin1red}) as one departs from the perturbative limit of small electric fields. At $\nu = {2}/{3}$ filling and $h=0$, we found that the system forms AB dimers for $J > \sqrt{2} \, t$. At small $h$, we can now make use of the effective description provided by Eq.~\eqref{eq:effham}. For $\ell_{\rm AB}=2$, the meson density is fixed to $\nu_{\rm m}(2) = {1}/{3} = \bar{\nu}_{\rm m}(2)$, and our effective model in the regime of low quantum fluctuations predicts  that we are in the critical regime $\delta_{\rm c}=1$, as $t_{\ell_{\rm AB}=2}= {h^2 t^2}/{J}{(J^2-2 \, t^2)}$. The groundstate will then be a LL of mesons, which propagate with an anomalous dimension reflected by the exponent of the algebraic decay of the composite boson gauge-invariant correlators $\langle b^\dagger_i b_{i+r}\rangle$ - see Fig.\ref{fig:dimercorr}. 

Up to now, we have observed that interference effects induced by the AB effect on the elementary plaquette loops result in a caging that locks particles in pairs inside specific partitions of the chain. We shall now show how this translates into a strong meson confinement even in the absence of strong electric-field lines connecting the charges. Remarkably, the emergent dynamics of the mesons bound inside the AB cages at small $h/t$ and $h/J$, also emerges for large electric field strength $h$, where the charges  are bound in meson pairs due to a  linearly-increasing confining potential. The effective model in this regime is still given by Eq.~~\eqref{eq:effham} with a 2-site constraint, and we find that the parameters are again fixed to the critical value $\delta_{\rm c} = 1$, analogously to what happens for the standard $\mathbb{Z}_2$ chain in this limit of strong electric fields~\cite{borla2020confined,kebric2021confinement}.

We now show that this tight confinement actually occurs  also when $J$ is comparable to $h$. For  $t=0$, the ground state manifold is obtained by diagonalizing the local Hamiltonian \eqref{eq:spin1red} $H_{\text{local}} = \sum_i h \, S^x_{i_\ell} + J (S^z_{i_\ell})^2$. The three  eigenvalues are $\epsilon_{\Uparrow} = {J}/{2} + \sqrt{{J^2}/{4} + h^2}$,  $\epsilon_0 = J$, and $\epsilon_{\Downarrow} = {J}/{2} - \sqrt{{J^2}/{4} + h^2}$, with the corresponding eigenstates  obtained as combinations of 0 and $\pi$-flux states $|\epsilon_{\Uparrow}\rangle = |\Psi^+\rangle + ({J+\sqrt{J^2+4h^2}}) |\Phi^+\rangle/{2h}$, $|\epsilon_0\rangle=|\Phi^-\rangle$ and  $|\epsilon_{\Downarrow}\rangle=|\Psi^+\rangle+(J-\sqrt{J^2+4h^2}) |\Phi^+\rangle/{2h}$. By gauge invariance, the ground state at $t/h = 0$ will be made up of tightly-bound pairs of particles separated by an $\ket{\epsilon_0}$ link, while all the other links shall arrange in the lowest energy state $\ket{\epsilon_{\Downarrow}}$. Along the same lines of the previous discussion, we can obtain a bosonic effective theory for composite mesons in which  $\tilde{b}_i = a_i \, a_{i+1} (\ket{\epsilon_{\Downarrow}} \bra{\epsilon_0})_{i_\ell}$ annihilate a dimer at site $i$. The dynamics of these mesons is then recovered at second order in a Schrieffer-Wolff perturbative expansion~\cite{schrieffer1966relation,bravyi2011schrieffer}. In this formalism, the mesons can move by second-order processes in which one of the charges tunnels such that the electric-field string connecting the pair of charges gets virtually stretched, and then it's subsequently compressed when the remaining charge of the dimer also tunnels in the same direction.
\begin{figure}
    \centering
    \includegraphics[width=\linewidth]{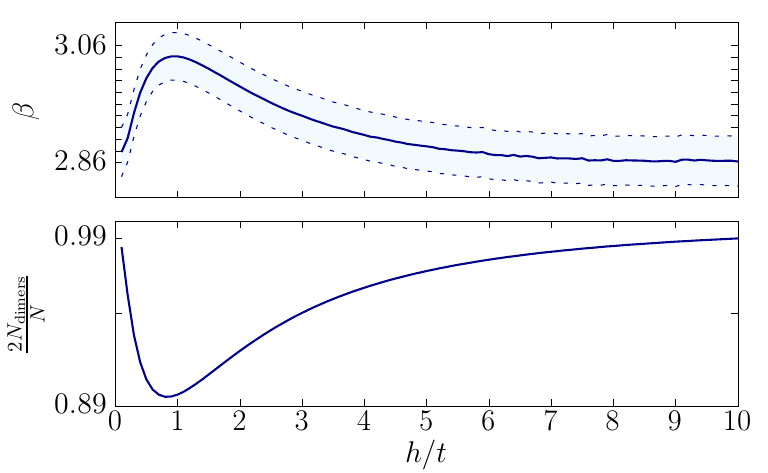}
    \caption{\textit{Algebraic decay of dimer correlators}. Upper panel: the exponent $\beta$ of the decay of dimer correlators $\langle b^\dagger_i b^{\phantom{\dagger}}_j \rangle \sim \abs{i-j}^{-\beta}$ is obtained from a best-fit analysis (solid blue line), together with the associated errors $\epsilon_\beta$ (shaded blue area between the dashed lines dermarking $\beta \pm \epsilon_\beta$). It is shown that dimers display almost the same algebraic decay throughout the parameter regime between the two limiting cases of small and large electric field. The anomalous dimension $\eta = \beta - 1$ of such composite mesons, is seen to vary with the ratio $h/t$. Lower panel: we plot the fraction of the total number of particles that form dimers. The ground-state results to be densely populated by dimers, even for intermediate values of $h/t$. The size of the chain and the filling fraction are set to $L=120$ and $\nu=2/3$ respectively.}
    \label{fig:anomalous}
\end{figure}

The effective Hamiltonian is again that of Eq.~\eqref{eq:effham} setting $\ell_{\rm AB} = 2$, albeit in this limit the 2-site constraint refers to the extent of the electric-field string instead of the AB cage. Notably, we find again $\delta_{\rm c}=V_{2}/{2 t_{2}} = 1$ - see Supplemental Material. Therefore, following the logic discussed previously for small $h/t$ and $h/J$, we can conclude that  the system  lies at the boundary between the LL and a gapped Mott insulating phase for any values of $J/t$, provided that we are in the strong electric field limit $t\ll h$. 

In spite of the different mechanisms behind confinement, as reflected  in the definitions of the bosonic operators in the two cases, the  two meson configurations share correlation functions $\langle b^ \dagger_i b_{i+r} \rangle$ and $\langle \tilde{b}^\dagger_i \tilde{b}_{i+r} \rangle$ that display the same algebraic decay - see Fig.~\ref{fig:dimercorr}. We can thus infer that tightly confined mesons produced either by AB interference at small $h/t$ and $h/J$, or by a strong electric field, move with the same anomalous dimension. Indeed, the two regimes are adiabatically connected, as can be inferred by inspection of the spectral gap, which remains closed from $J>\sqrt{2}t$ and small $h/J$,$h/t$ to the regime of strong electric field, for any values of $J/t$ -- see SM, so that the system remains in a LL phase. Whilst dimers fully populate the ground-state only in the limiting regimes of large and small electric field, we can extend the definition of dimer operators $\tilde{b}_i = a_i \, a_{i+1} (\ket{\epsilon_{\Downarrow}} \bra{\epsilon_0})_{i_\ell}$ to intermediate values of $h/t$, as the state $\epsilon_{\Downarrow}$ varies with this ratio. Note that, with changing $h/t$, both the limiting cases are recovered. Inspection of the one-body dimer correlators shows that these mesons display almost the same algebraic decay throughout the whole parameter regimes comprised between the two analytical limits, with an anomalous dimension that varies with $h/t$ - see Fig.\ref{fig:anomalous}. Remarkably, dimers are moreover shown to densely populate the ground-state, also at intermediate values of the electric field to hopping amplitude ratio - see Fig.\ref{fig:anomalous}.

\subsection{$\mathbb{Z}_3$ Mott insulator of mesons locked into AB trimers}
\label{sub:mott}
\begin{figure}
    \centering
    \includegraphics[width =\linewidth]{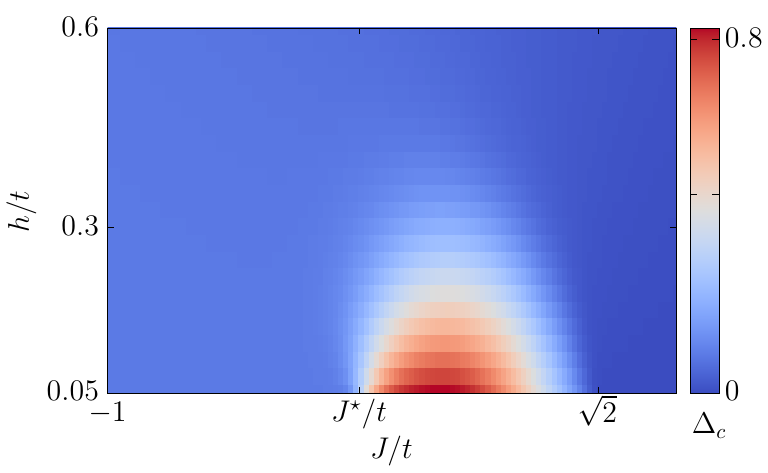}
    \caption{\textit{Charge gap $\Delta_c$}. A charge gap opens up at the transition between a LL and a $\mathbb{Z}_3$ Mott insulator around $J = \sqrt{2} \, t$, for $\nu = \frac{2}{3}$. The Mott phase, which at $h/t, h/J \ll 1$ spans the range $\frac{J^\star}{t} < \frac{J}{t} < \sqrt{2}$ extends in a lobe structure to finite values of the electric field strength $h$. The system size is set to $L=120$.}
    \label{fig:charge_gap}
\end{figure}
\begin{figure}
    \centering
    \includegraphics[width=\linewidth]{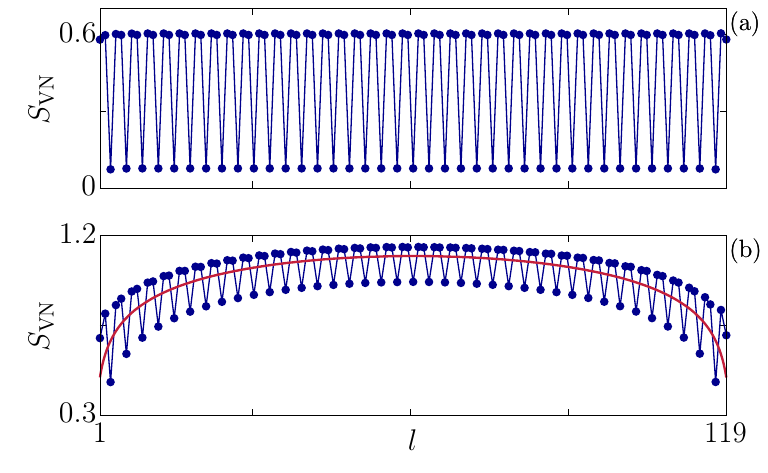}
    \caption{\textit{von Neumann Entanglement Entropy.} We plot the entanglement entropy (EE) for bipartitions of the lattice into two subchains of length $l$ and $L-l$, where $L=120$ and the filling is fixed to $\nu = \frac{2}{3}$. The electric field strength is fixed to the value $h/t = 0.2$. (a) Trimers Mott regime at $J/t = 0.8$: the ground state is quasi-factorized in the tensor product of trimer ground states. (b) Dimer LL regime at $J/t = 1.6$: the EE follows the CFT behaviour (red solid line) $S_{\text{CFT}}(l) = \frac{c}{6} \log(\frac{2L}{\pi} \sin(\frac{\pi l}{L})) + \tilde{c}$, with central charge $c=1$. By fitting the data we find $\tilde{c} \sim 0.37$.}
    \label{fig:EE}
\end{figure}
In the previous section, we have shown that tightly confined mesons arise either as a result of AB interference at small $h/t$ and $h/J$, or due to a strong confining interaction at large $h/t$. At $\nu={2}/{3}$ filling, the ground state in these two regimes lies at the critical point $\delta_c = 1$ of the effective perturbative model \eqref{eq:effham}, between a LL of $\ell_{\rm AB} = 2$ cages and a Mott insulating phase in which such dimers arrange themselves in a configuration that breaks translational symmetry to its $\mathbb{Z}_3$ subgroup. We find that a $\mathbb{Z}_3$ Mott insulating phase can actually be  stabilized in our $\mathbb{Z}_2$-loop chain by lowering the magnetic coupling $J$. Indeed, in the grand canonical ensemble, we  found an incompressible phase in which AB trimers completely cover the chain when $(J^\star < J < \sqrt{2} \, t)$. When quantum fluctuation are introduced  $h\neq0$, and in the regime $J \approx\sqrt{2} \, t$, the size of the cages  fluctuates and $\ell_{\rm AB} = 2$ and $\ell_{\rm AB} = 3$ cages can become energetically degenerate. We find that, when lowering the magnetic coupling further, a transition to a different configuration can take place, as AB trimers become  energetically favourable, each of them hosting two charges bound in a meson that now extends to the three sites. This forms a $\mathbb{Z}_3$ Mott insulator as reflected by the vanishing compressibility $\kappa=0$ reported earlier, as well as is reflected by the opening of a charge gap $\Delta_c = \frac{1}{2} [E(N+2) - 2 \, E(N) + E(N-2)]$. Here $E(N)$ is the ground-state energy with $N$ particles and the choice $N \pm 2$ is needed to fulfill the even parity requirement in the chosen gauge sector. The numerical results are obtained through DMRG and show that this Mott phase extends to finite values of the electric field strength $h$. Indeed, both the charge gap and the spectral gap - see Fig.~\ref{fig:charge_gap} and SM, remain open at finite $h$, so that the ground-state of the system at $\nu={2}/{3}$ filling is in a $\mathbb{Z}_3$ Mott insulating phase within values of the magnetic coupling comprised between $J^\star$ and $\sqrt{2}t$. Outside of this region, the system falls back into the LL of mesons.
 
To corroborate the previous analysis, we consider a bipartition of the chain in two subchains of length $l$ and $L-l$ and we calculate the von Neumann entanglement entropy (EE) of the bipartition as a function of $l$ - see Fig.~\ref{fig:EE}. We find that, in the $\mathbb{Z}_3$ Mott insulator, the EE nearly vanishes when the bipartition does not intersect any of the sites enclosed in the AB trimer cage, which is when the length of the bipartition $l$ is an integer multiple of $3$. This behaviour reflects a  quasi-factorized ground state at these specific bipartitions. In the dimer regime, in contrast, the EE displays the conformal field theory (CFT) behaviour, which is consistent with the expectation for a LL with central charge $c=1$ in a finite-size system \cite{calabrese2009entanglement}.

Let us close this section by noting that  in the standard $\mathbb{Z}_2$ chain, in order to find these Mott insulating phases, one needs to introduce additional density-density Hubbard-type interactions between charges at neighboring sites, which can favour a charge-density-wave pattern and effectively lead to  $\delta > 1$ favouring the Mott insulator~\cite{kebric2021confinement}. In our case, the origin  of the Mott insulator is very different: it results from the interplay of the charge dynamics and the dynamical AB caging. We also note that a similar analysis applies to the half-filled  case $\nu={1}/{2}$, as the system can transition from a LL of trimer cages, as described by the effective model \eqref{eq:effham} with $\delta < 1$, to a Mott insulator of fourmer cages at the full-filling meson density $\tilde{\nu}_b{(4)} = {1}/{4}$ - see Supplemental Material.
\end{subsection}
\end{section}

\begin{section}{Conclusions and outlook}
\label{sec:conclusions}
In this paper, we have introduced a minimal quasi-1d geometry in which $\mathbb{Z}_2$ gauge fields can lead to the formation of dynamical Aharonov-Bohm cages. While bearing some resemblance to the 'static cases' provided by an external magnetic field~\cite{vidal2000interaction,doucot2002pairing,PhysRevLett.107.150501,cartwright2018rhombi}, the physics of  dynamical caging  in our lattice gauge theory is  substantially different. Indeed, our  caging arises from the competition of matter and gauge-field dynamics, and is not a consequence of the appearance of flat bands in the  single-particle energy spectrum, as it is the  case in quantum Hall type systems~\cite{vidal2000interaction}. We have shown that these cages self-arrange in periodic structures, separated by $\pi$ flux loops, providing the analog of the two-dimensional visons, thus breaking the  translational symmetry of the system. Pairs of $\mathbb{Z}_2$ charges get effectively confined inside the cages, forming tightly-confined mesons. In contrast to the standard $\mathbb{Z}_2$ chain, matter is  tightly confined even at a vanishingly-small value of the electric field  $h$. The  interplay between  caging and confinement results in the different ways in which the chain fragments into sub-chains of different sizes, which in turn depends on the ratio of the effective magnetic flux and tunneling strengths $\displaystyle{{J}/{t}}$. 
We note that, while at $h=0$ confinement arises from the aforementioned chain's fragmentation (see section \ref{sub:h_0}), quantum fluctuations of the magnetic fluxes induced by a non-vanishing electric field $h>0$ actually restore the tunneling among disconnected clusters;  this effect provides  the key allowing mesons to propagate in the system -- see section \ref{sub:small_h}. For small values of $h$, such dynamics is shown to be integrable, and leads to a gapless Luttinger Liquid (LL) of mesons -- see Eq.~\eqref{eq:effham} and the discussion below. The LL behaviour at small $h/t$ and $h/J$, is shown to adiabatically extend to finite values of the electric field strength, up to the limit of large $h/t$. In the latter case, particles are shown to be also tightly-confined into mesons with a short electric-field line that involves a   single link, similar to the case of the standard $\mathbb{Z}_2$ chain. In this regime, confinement does not arise from a  destructive interference induced by the gauge flux, but it is due to the linearly-growing attractive interactions between particles separated by electric field strings. Remarkably, in spite of these different origin of confinement, the dimerized regimes at weak and strong electric field $h$ showcase the same quasi-long-range behaviour of the dimer correlators -- see Fig.~\ref{fig:dimercorr}. We thus find a novel mechanism for the tight confinement of $\mathbb{Z}_2$ charges where both electric-field penalties and interference effects contribute,  this effect persisting also for intermediate values of $h/t$ -- see Fig.\ref{fig:anomalous}. 

For specific values of the filling fraction, we discover that quantum phase transitions to incompressible Mott-insulating phases occur -- see Sec.\ref{sub:mott}. These transitions, which result in the opening of a charge gap (see Fig.~\ref{fig:charge_gap}), arise  from  quantum fluctuations triggered by the electric field,  that aim at modifying the size of the AB cages $\ell_{\rm AB}$. At a critical value, one finds quasi-degenerate configurations of the clusters characterized by either $\ell_{\rm AB}$ or $\ell_{\rm AB}+1$ sites, with the largest cage  covering the whole lattice. As the magnetic flux strength is lowered below its critical value, larger cages are favoured and stabilize a Mott insulating configuration with broken translational symmetry. These phases are characterized by a vanishing compressibility $\kappa$: changing the chemical potential $\mu$ at fixed ${{J}/{t}}$ has no effect, until a new commensurate covering of the lattice with cages of different sizes becomes energetically favourable -- see Fig.~\ref{fig:mott_lobes}.

In future theoretical studies,  different constraints on the particle dynamics (Gauss laws) could be explored, as they can lead to different cages' filling. Finite temperature effects could also be addressed, as thermal fluctuations are expected to compete with the discussed caging phenomenon. 

Our work identifies AB caging as a new driving force that can affect confinement, and it would be very ineteresting to explore it in fully two-dimensional models. Based on the scenario emerged in the present study, our dynamical caging might define a new form of string-net theories\cite{buerschaper2009mapping,levin2005string,wen2017zoo}.   

Finally, we note that experimental realizations of the phenomenology described in this paper are in line with current proposals for the quantum simulation of $\mathbb{Z}_2$ gauge theories, f.i. through Rydberg atoms or trapped ions\cite{Bazvan2024synthetic,domanti2024floquet}.  In the latter platform, real-time dynamics and the interplay of AB interference for a single $\mathbb{Z}_2$ loop have been demonstrated in recent experiments ~\cite{oxford_z2_gauge}. 

\end{section}

\acknowledgements
We thank J. Polo, E.Tirrito, F. Perciavalle, P. Kitson, L. Chirolli and W.J.Chetcuti for  useful discussions.
A.B.  thanks  O. Băzăvan and S. Saner for discussion related to the trapped-ion implementation of the single-loop $\mathbb{Z}_2$ model, and the role of the Bell pairs with a well-defined magnetic flux for reaching gauge invariance. A.B. acknowledges support from PID2021-127726NB-
I00 (MCIU/AEI/FEDER, UE), from the Grant IFT Centro
de Excelencia Severo Ochoa CEX2020-001007-S, funded
by MCIN/AEI/10.13039/501100011033, from the CSIC Research Platform on Quantum Technologies PTI-001, and from
the European Union’s Horizon Europe research and innovation programme under grant agreement No 101114305
(“MILLENION-SGA1” EU Project).

\bibliography{bibliography}

\onecolumngrid
\newpage

\pagebreak
\widetext
\begin{center}
\textbf{\large Supplemental material for 'Dynamical Aharonov-Bohm cages and tight meson confinement in a $\mathbb{Z}_2$-loop gauge theory'}
\end{center}


\setcounter{equation}{0}

\setcounter{figure}{0}

\setcounter{section}{0}

\setcounter{table}{0}

\setcounter{page}{1}

\makeatletter

\renewcommand{\theequation}{S\arabic{equation}}

\renewcommand{\thefigure}{S\arabic{figure}}

\renewcommand{\bibnumfmt}[1]{[S#1]}

\renewcommand{\citenumfont}[1]{S#1}

\renewcommand{\thesection}{\Alph{section}}

In this Supplemental Material, we provide details that support the discussion of the main text. We are concerned with the loop-chain $\mathbb{Z}_2$ lattice gauge theory, which we introduced in Eq.(1) of the main text as a modified $\mathbb{Z}_2$ chain \cite{borla2020confined,kebric2021confinement}, with an additional Ising spin on each bond of a one-dimensional lattice. Particles are hardcore bosons that can hop along the chain between neighboring sites, following either of two different paths, each corresponding to one of the Ising spins on the links. Such paths close in a loop geometry between every pair of neighboring sites, so that we call the system a "loop chain". In section IIB of the main text we have discussed that the system locally conserves the total spin squared $S^2_{i_\ell}$ of each loop, with $\boldsymbol{S}_{i_\ell} = (\boldsymbol{\sigma}_{1,i_\ell} + \boldsymbol{\sigma}_{2,i_\ell})/2$. Ignoring singlet eigenstates of $S^2_{i_\ell}$, which correspond to the eigenvalues $s_{i_\ell} = 0$ and that are dark-states completely decoupled from the dynamics of the system, the Hamiltonian of the model is rewritten as
\begin{equation}
    \label{app:eqn:spin1red}
    \mathcal{H} = t \sum_i (a_i^\dagger S^z_{i_\ell} a_{i+1} + {\rm H.c.}) + h \sum_i S^x_{i_\ell} + J \sum_i (S^z_{i_\ell} )^2 \, ,
\end{equation}

\ni whose bulk gauge-symmetry generators are
\begin{equation}
    \label{app:eqn:generators}
    G_i = P^x_{i_\ell-1} (-1)^{n_i} P^x_{i_\ell} \, ,
\end{equation}

\ni where $P^x = 2 (S^x)^2 -1$. At the edges of a chain terminating with sites, the gauge-symmetry generators are deformed to $G_1 = (-1)^{n_1} P^x_{1_\ell}$ and $G_L = P^x_{L_\ell -1} (-1)^{n_L}$.

The Supplemental Material is organized as follows. In section \ref{app:sec:spin1_map}, matter degrees of freedom are integrated out by making use of the Gauss' law in the neutral gauge sector $G_i=1$, $\forall i$, in favour of a purely spin-$1$ model. In section \ref{app:sec:dmrg}, a rewriting of the model is adopted that allows to enforce all of its symmetries within DMRG simulations. The limit of vanishing electric field $h=0$ is addressed in section \ref{app:sec:zero_field}, where we provide details on the formation of the dynamical Aharonov-Bohm(AB) cages discussed in the article, both with varying particle number, as controlled by a chemical potential term $-\mu \sum_i n_i$ and at fixed values of the filling fraction. The perturbative regimes of small and large electric field are ultimately addressed in sections \ref{app:sec:small_h} and \ref{app:sec:large_h}, respectively.

For later convenience, we will consider a Jordan-Wigner transformation $c_i = \prod_{j<i} (-1)^{a^\dagger_j a_j} a_i$ which, for open-boundary conditions, just amounts to the replacement of the hardcore boson operators $a$ ($a^\dagger$) with their fermionic counterparts $c$ ($c^\dagger$), as all the terms in the Hamiltonian consist of local or nearest-neighbours combinations of matter operators. We will then refer equivalently to the fermionic model
\begin{equation}
    \label{app:eqn:fermionic_spin1red}
    \mathcal{H} = t \sum_i  (c_i^\dagger S^z_{i_\ell} c_{i+1} + H.c.) + h \sum_i S^x_{i_\ell} + J \sum_i (S^z_{i_\ell})^2 - \mu \sum_i n_i \, .
\end{equation}

\section{Mapping to a spin-$1$ model}
\label{app:sec:spin1_map}
In this section, we demonstrate that the dynamics of Eq.\eqref{app:eqn:spin1red} can be recast in terms of a spin-$1$ model, in which matter has been integrated out by using the Gauss-law, along the same lines as for the standard $\mathbb{Z}_2$ chain \cite{borla2020confined,borla2021gauging}. We will start from the fermionic representation of the model, Eq.\eqref{app:eqn:fermionic_spin1red}. In the bulk of a chain of size $L$ ending with sites, we can make use of the Gauss-law in the neutral gauge sector $G_i = 1$ to relate the matter density at site $i$ to the spin degrees of freedom living on the neighboring bonds $i_\ell$ and $i_\ell-1$:
\begin{equation}
\label{app:eqn:domainwalls}
    n_i = \frac{1}{2} \Big(1-P^x_{i_\ell-1} P^x_{i_\ell}\Big) \, .
\end{equation}

\ni As is evident from the previous equation, in the spin-1 language particles become domain walls separating regions of ferromagnetic ordering for the $\mathbb{Z}_2$ valued eigenstates of $P^x$.
Introducing the Majorana combinations
\begin{align}
    &\gamma_i = c_i^\dagger + c_i  \\
    &\tilde{\gamma}_i = \text{i} (c_i^\dagger - c_i) 
\end{align}

\ni which satisfy $\{\gamma_i,\tilde{\gamma}_j\} = 0$ and $\{\gamma_i,\gamma_j\} = \{\tilde{\gamma}_i,\tilde{\gamma}_j\} = 2 \delta_{i,j}$, the Gauss law is rewritten as $G_i = P^x_{i-\frac{1}{2}} \, \text{i} \tilde{\gamma}_i \gamma_i \, P^x_{i+\frac{1}{2}} = 1$, where $\text{i} \tilde{\gamma}_i \gamma_i = (-1)^{n_i}$. The Hamiltonian, recast in terms of the Majorana operators, reads (except for a constant term):
\begin{equation}
\label{app:eqn:majorana_ham}
    \mathcal{H} = \frac{\text{i} \, t}{4} \sum_i (\gamma_i \, S^z_{i_\ell} \,\tilde{\gamma}_{i+1} - \tilde{\gamma}_i \, S^z_{i_\ell} \, \gamma_{i+1} - H.c.) + h \sum_i S^x_{i_\ell} + J \sum_i (S^z_{i_\ell})^2 + \text{i} \frac{\mu}{2} \sum_i \tilde{\gamma}_i \gamma_i \, .
\end{equation}

Spin-$1$ operators can be introduced as follows
\begin{equation}
\label{app:eqn:spin1operators}
\begin{aligned}
    &X_{i_\ell} = S^x_{i_\ell} \\
    &Y_{i_\ell} = - \text{i} \tilde{\gamma}_i \, S^y_{i_\ell} \, \gamma_{i+1} \\
    &Z_{i_\ell} = - \text{i} \tilde{\gamma}_i \, S^z_{i_\ell} \, \gamma_{i+1} \\
\end{aligned}
\end{equation}

\ni that satisfy the commutation relation of SU(2). All the terms in Eq.\eqref{app:eqn:majorana_ham} can then be rewritten in terms of these spin operators. We will make use of the following identities:
\begin{align}
\label{app:eqn:id1}
    & \text{i} \tilde{\gamma}_i \gamma_i = P^X_{i_\ell-1} P^X_{i_\ell} \\
\label{app:eqn:id2}
    &(S_z^2)_{i_\ell} = Z^2_{i_\ell} \\
\label{app:eqn:id3}
    & \text{i} \gamma_i \, S^z_{i_\ell} \, \tilde{\gamma}_{i+1} = -P^X_{i_\ell-1} Z_{i_\ell} P^X_{i_\ell+1}
\end{align}

\ni where the first is a rewriting of the Gauss-law in the neutral gauge sector and $P^x$ has been directly mapped to $P^x \to P^X = 2 X^2 - 1$, the second is obtained by making use of the fact that $\gamma^2 = \tilde{\gamma}^2 = 1$ (as is inferred by the anti-commutation relation of the Majorana operators), while the third is obtained introducing successive identities through the Gauss-law $1 = P^x_{i_\ell-1} \, \text{i} \tilde{\gamma}_i \gamma_i \, P^x_{i_\ell}$. In detail, we have:
\begin{align*}
    \text{i} \gamma_i \, S^z_{i_\ell} \, \tilde{\gamma}_{i+1} &= \text{i} \gamma_i (P^x_{i_\ell-1} \, \text{i} \tilde{\gamma}_i \gamma_i \, P^x_{i_\ell}) S^z_{i_\ell} (P^x_{i_\ell} \, \text{i} \tilde{\gamma}_{i+1} \gamma_{i+1} \, P^x_{i_\ell+1}) \tilde{\gamma}_{i+1} = - \text{i} \tilde{\gamma}_i \, P^x_{i_\ell-1} (P^x_{i_\ell} S^z_{i_\ell} P^x_{i_\ell})P^x_{i_\ell+1} \gamma_{i+1} \\
    &= - P^x_{i_\ell-1} (- \text{i} \tilde{\gamma}_i \, S^z_{i_\ell} \, \gamma_{i+1}) P^x_{i_\ell+1} = -P^X_{i_\ell-1} Z_{i_\ell} P^X_{i_\ell+1} \, ,
\end{align*}

\ni where the first equality is obtained by introducing the Gauss-law twice, the second comes from the anti-commutation relations of the Majorana operators, the third is a consequence of $P^x S^z P^x = - S^z$, while the last is achieved making use of Eq.\eqref{app:eqn:spin1operators}. Eventually, neglecting an additive constant, we can rewrite Eq.\eqref{app:eqn:majorana_ham} in the bulk of the chain as a spin-$1$ model on the dual lattice $(i_\ell \to i+1)$:
\begin{equation}
    \label{app:eqn:spin1_model}
    \mathcal{H} = \sum_{i=2}^{L-2} \left\{ \frac{t}{2} \Big(1-P^X_{i-1} P^X_{i+1}\Big) Z_{i} + h X_i + J Z_i^2 + \frac{\mu}{2} P^X_{i-1}P^X_i \right\} \, .
\end{equation}

\ni The mapping described above solely applies to the bulk of the chain, as we have made use of the Gauss-law for the bulk. At the edges of a chain terminating with sites, we can follow an analogous approach by using the Gauss-law at the first and last sites of the lattice, namely $G_1 = \text{i} \tilde{\gamma}_1 \gamma_1 P^x_{1_\ell} = 1$ and $G_L = P^x_{L_\ell-1} \text{i} \tilde{\gamma}_L \gamma_L = 1$. Proceeding in the same way and neglecting an irrelevant constant term, we can write the contribution of the edges as
\begin{equation}
    \mathcal{H}_{\text{edges}} = t \left[ Z_1 (1-X_2^2) + (1-X^2_{L-2}) Z_{L-1} \right] + \frac{\mu}{2} \left(P^X_1 + P^X_{L-1} \right) + h \left(X_1 + X_{L-1}\right) + J \left(Z_1^2 + Z_{L-1}^2\right) \, .
\end{equation}

\section{DMRG encoding}
\label{app:sec:dmrg}
The main results in the article have been obtained using MPS DMRG algorithms \cite{schollwock2011the}, through the ITensors library for Julia \cite{fishman2022the}. We have decided not to apply DMRG directly to the model \eqref{app:eqn:spin1red}, for reasons that we will explain in the following, but to resort to a mapping to another model, whose dynamics reduces to that of Eq.\eqref{app:eqn:spin1red} when certain conditions are fulfilled. Indeed, the Hamiltonian in Eq.\eqref{app:eqn:spin1red} consists of both matter and gauge degrees of freedom living on the sites and on the links of the lattice, respectively. A state of the system will then be encoded in a MPS, which will carry physical indices (site indices) corresponding to the local Hilbert spaces of the hardcore bosons, of dimension $d=2$, and of the spin-$1$ gauge-fields, of dimension $d=3$. When conservation laws need to be imposed, that arise from certain symmetries of the system, one can resort to quantum-number conserving algorithms which require that the local density of the conserved symmetry operator has support on the individual sites of the MPS. In the case of our interest, this is true for the particle number conservation arising from the $U(1)$ symmetry of model \eqref{app:eqn:spin1red}, corresponding to the local density $n_i$, with $n_i$ being the on-site occupation number of the hardcore bosons. The extensively many constraints imposed by the Gauss' law, on the other hand, each involve operators acting on three neighboring sites of the MPS, one for matter and two for the neighboring gauge fields. Restriction to a certain gauge sector can thus be enforced by penalizing energetically the overlap to states which are unphysical. This is achieved through Hamiltonian penalty terms that, for the neutral gauge sector $G_i = 1$, are of the form $-\lambda \sum_i G_i$. Such terms suppress unwanted contributions from other gauge-sectors in the low-energy regime, but are hard to control \cite{Felser2020two}, as large penalties can cause the DMRG algorithm to get stuck into unwanted minima of the energy. A possible solution to this issue would be to resort to the matter elimination described in the previous section \ref{app:sec:spin1_map}, in which the constraints of the Gauss law are automatically enforced. In this case, one would deal with a spin-$1$ chain, so that an MPS describing a state of the system will carry physical indices for each spin along the chain. Such an approach comes with a similar issue when we want to make use of the global $U(1)$ symmetry of the model to restrict to subspaces of fixed number of particles. Indeed, the particle density translates into domain wall operators that extend over two-sites -- see Eq.\eqref{app:eqn:domainwalls}, so that fixing the number of particles will require a fine-tuning of the chemical potential term $\mu$ -- see Eq.\eqref{app:eqn:spin1_model}.

As discussed in the main text -- see Sec.IIC of the main text and the figures therein, we propose a solution to these issues by resorting to a mapping to a different model. Our method is summarized as follows
\begin{enumerate}
    \item The links of the chain are broken into halves and each of these half-links is attributed to its nearest neighbouring site. Super-sites are then defined that include the original matter sites and the two adjacent half-links. 
    \item Each original spin-1 operator $S^\alpha_{i_\ell}$ is replaced by an Ising spin $\sigma^\alpha_i$ living on the half-link at the right of the super-site at position $i$ and a spin-1 operator $S^\alpha_{i+1}$ living in the left half-link of the super-site at $i+1$.
    \item On this modified chain of super-sites, we introduce the Hamiltonian
    \begin{equation}
    \label{app:eqn:dmrg_mapped}
        \tilde{H} = t\sum_i  (a_i^\dagger \sigma_i^z S_{i+1}^z a^{\phantom{\dagger}}_{i+1} + {\rm H.c.})  + h\sum_i S_i^x + J \,\sum_i (S_i^z)^2 \, ,
    \end{equation}
    which has a $\mathbb{Z}_2$ gauge-symmetry generated by operators defined at each super-site, namely $\tilde{G}_i = P^x_i (-1)^{n_i} \sigma^x_i$, $\tilde{G}_1 = (-1)^{n_1} \sigma_1^x$ and $\tilde{G}_L = P_L^x (-1)^{n_L}$. An MPS describing a state of the system will now carry physical indices corresponding to each super-site, so that the local Hilbert space for each site in the bulk will have dimension $d_i=12$. At the edges, the super-site at the left will consist of a matter site and an Ising spin degree of freedom, so that the corresponding Hilbert space will have size $d_1=4$. The local space of the rightmost super-site will instead be the tensor product of the Hilbert spaces of an hardcore boson and a spin-1 degree of freedom, having dimension $d_L=6$. The dynamics of the spin-1 reduction of the loop-chain Hamiltonian, Eq.\eqref{app:eqn:spin1red}, is recovered by imposing the conditions that
    \begin{equation}
    \label{app:eqn:dmrg_constraints}
        \sigma^x_i = P^x_{i+1} \, .
    \end{equation}
    The main idea is that the Ising spins are introduced to act as trackers of the state of $P^x$ on the neighboring super-site, in order to recover the same form of the Gauss' law when the constraints in Eq.\eqref{app:eqn:dmrg_constraints} are fulfilled, as they imply $\tilde{G}_i = G_i$. As a consequence, $\sigma^x_i$ and $P^x_{i+1}$ need to change simultaneously when particles tunnel, thus motivating the form of the hopping term in the introduced model \eqref{app:eqn:dmrg_mapped}. To enforce Eq.\eqref{app:eqn:dmrg_constraints} we explicitly apply the projectors $\mathcal{P} = \prod_i (1+\sigma_i^x P^x_{i+1})/2$ to the Hamiltonian $\tilde{H}$, so that we will work with $\mathcal{P} \tilde{H} \mathcal{P}$.
    \item We will restrict ourselves to the gauge-sector determined by the condition that $\tilde{G}_i = 1$ $\forall i$, which can be enforced by eliminating the "unphysical" states not belonging to the chosen gauge sector from the local Hilbert space of each super-site. Each physical index of the MPS will now span a space of dimension $d_i=6$ in the bulk and $d_1 = 2$, $d_L = 3$ at the edges. With this choice of gauge-sector and with the constraints in Eq.\eqref{app:eqn:dmrg_constraints} we recover the dynamics of Eq.\eqref{app:eqn:spin1red} with the Gauss' law $G_i = 1$ $\forall i$. 
    \item While the effect of $\mathcal{P}$ is to give dynamics only to those states that satisfy Eq.\eqref{app:eqn:dmrg_constraints}, a random MPS will still have projections over all the other unwanted states, that give a zero contribution to the energy. Since the constraints involve operators that have support on neighboring super-sites, an energetic penalty would then be needed to ensure that the DMRG procedure converges to the correct ground-state. Nevertheless, we construct a global symmetry operator $O = \sum_i O_i$, whose density $O_i$ only involves single-site operators and is such that a specific symmetry sector does not contain any of the unwanted states. Taking $O_i = 2^{i-2} P_i^x - 2^{i-1} \sigma_i^x$, we can write
\begin{equation}
\label{app:eqn:dmrg_global}
    O = \sum_{i=1}^{L-1} O_i = \sum_{i=1}^{L-1} 2^{i-1} (P_{i+1}^x - \sigma_i^x) \, .
\end{equation}

 \ni The operator $O$ commutes with
 $\mathcal{P}^\dagger \tilde{H} \mathcal{P}$ by construction and $\gamma_i= P_{i+1}^x - \sigma_i^x$ has eigenvalues $\{-2,0,2\}$. Note that $\gamma_i = 0$ corresponds to the condition \eqref{app:eqn:dmrg_constraints}. We shall prove that, restricting to the subspace in which $O=0$, all the aforementioned constraints are automatically enforced, as none of the states violating the conditions \eqref{app:eqn:dmrg_constraints} belongs to this symmetry sector. The condition that $O=0$ translates into
 \begin{equation}
     \sum_{i=1}^{L-1} 2^{i-1} \gamma_i = \gamma_1 + 2\gamma_2 + \dots + 2^{L-2} \gamma_{L-1} = 0 \, .
 \end{equation}
 If we assume that some of the $\gamma_i$ eigenvalues are different from zero, for $i$ in an {ordered set} $\{i_1,\dots,i_n\}$, than we can rewrite them as $\gamma_i = 2 s_i$, with $s_i \in \{-1,1\}$. Notice that this last statement comes from the fact that the non-zero eigenvalues of $\gamma_i$ are $\pm 2$. The condition $O=0$ then becomes
 \begin{equation}
     s_{i_1} 2^{i_1} + \dots s_{i_n} 2^{i_n} = 2^{i_1} (s_{i_1} + s_{i_2} 2^{i_2 - i_1} + \dots s_{i_n} 2^{i_n - i_1}) = 0
 \end{equation}
 which is never fulfilled, since it would require that $s_{i_1} = -(s_{i_2} 2^{i_2 - i_1} + \dots s_{i_n} 2^{i_n - i_1})$, but the left hand side of the equation is an odd number, while the right hand side is even. By contradiction, we have then proved that, in the subsector $O=0$, all the $\gamma_i$ need to vanish and only states satisfying the constraints \eqref{app:eqn:dmrg_constraints} are present. 
\end{enumerate}

\section{Analysis at $h=0$}
\label{app:sec:zero_field}
In this section, we discuss the $h=0$ limit of Eq.\eqref{app:eqn:spin1red}. As presented in the main text, destructive interference effects determined by link states in $\pi$-flux configurations $|\Psi^+\rangle$ cause the lattice to be fragmented into AB cages. Such $\pi$-flux states are the analogue of visons of two-dimensional gauge theories. The AB cages enclose only $0$-flux states $\{ | \Phi^\pm \rangle \}$, but are bounded by visons, so that the particles cannot hop out of a cage due to destructive interference and are trapped inside. At $h=0$, the Wilson plaquette for each loop $W_{\bigcirc_{i_\ell}}= 2(S^z_{i_\ell})^2 -1$, whose expectation value $\langle W_{\bigcirc_{i_\ell}} \rangle$ represents the average gauge-flux threading the plaquette, is locally conserved. Therefore, different distributions of such vison states along the chain correspond to dynamically disconnected sectors of the Hilbert space, whose number, for a chain of $L$ sites, is given by all the possible partitions of the integer $L$ which are compatible with the even parity requirement imposed by the Gauss law. Indeed, each subchain will only be allowed to host zero or an even number of particles. The competition between the energetic contribution of the gauge flux $J$, the particles dynamics and their filling determines the most favourable configurations of cages in the ground-state of the system.

We shall begin our discussion considering the fermionized model \eqref{app:eqn:fermionic_spin1red}. Each independent cage of length $\ell_{\rm AB}$ will be described by the $\mathbb{Z}_2$-LGT Hamiltonian $H_{\ell_{\rm AB}} = J \, (\ell_{\rm AB}-1) + t \sum_{i = 1}^{\ell_{\rm AB}-1} (c_i^\dagger \sigma^z_{i_\ell} c_{i+1} + H.c.) - \mu \sum_{i=1}^{\ell_{\rm AB}} n_i$, where $\sigma^z$ denotes the reduction of $S^z$ to the space spanned by the zero-flux states $\{ |\Phi^\pm\rangle \}$ and the constant term corresponds to the gauge-flux contribution to the energy. Introducing new fermionic operators $f_i = \prod_{k<i} \sigma^z_{k_\ell} c_i$, the previous Hamiltonian is reduced to a model of free fermions hopping on a chain of length $\ell_{\rm AB}$, $H_{\ell_{\rm AB}} = J \, (\ell_{\rm AB}-1) + t \sum_{i=1}^{\ell_{\rm AB}-1} (f_i^\dagger f_{i+1} + H.c.) - \mu \sum_{i=1}^{\ell_{\rm AB}} n_i^f $, which is promptly diagonalized by the canonical transformation $f_n = \sqrt{\frac{2}{\ell_{\rm AB}+1}} \sum_{i=1}^{\ell_{\rm AB}} (-1)^i \sin{(\frac{n \pi i}{\ell_{\rm AB}+1})} f_i$, with $n=1, \dots, \ell_{\rm AB}$, yielding the spectrum $\epsilon_{\ell_{\rm AB}} = J \, (\ell_{\rm AB}-1) -\sum_{n} (2t \cos{(\frac{n \pi}{\ell_{\rm AB}+1})} + \mu) \, n_n^f$. 

To simplify the notation, in the following we shall indicate the size of the cages $\ell_{\rm AB}$ just as $\ell$. The single particle energy levels in a cluster of $\ell$ sites are given by (the hopping strength $t$ is set to $t=1$):
\begin{equation}
    \label{app:eqn:spe}
    \epsilon_{n,\ell} = -2\cos(\frac{n \pi}{\ell+1}) - \mu \, ,
\end{equation}
\ni where $n \in \{1,\dots,\ell\}$. Given a possible partitioning of the lattice into AB cages of length $\ell_i$, the lowest energy configuration is reached by filling up all the available single particle energy levels in each subchain, up to the largest even integer that satisfies $n \leq \frac{\ell + 1}{\pi} \arccos(-\frac{\mu}{2})$, corresponding to the condition $\epsilon_{n,\ell} \leq 0$. We shall denote this integer through a modified floor symbol $\lfloor \cdot \rfloor^\ast$, to remind ourselves about the even parity constraint. Introducing $\bar{N}(\ell,\mu) = \lfloor \frac{\ell + 1}{\pi} \arccos(-\frac{\mu}{2}) \rfloor$, the number of particles in each subchain is given by
\begin{equation}
    \label{app:eqn:maxinteger}
     N(\ell,\mu) = \Bigg\lfloor \frac{\ell + 1}{\pi} \arccos(-\frac{\mu}{2}) \Bigg\rfloor^\ast = \begin{cases}
         \bar{N}(\ell,\mu) \quad &\text{if $\bar{N}$ is even} \\
         \bar{N}(\ell,\mu) + 1 \quad &\text{if $\bar{N}$ is odd $\neq \ell$ and } \epsilon_{\bar{N},\ell} + \epsilon_{\bar{N}+1,\ell} < 0\\
         \bar{N}(\ell,\mu) - 1 \quad &\text{if $\bar{N}$ is odd and } \epsilon_{\bar{N},\ell} + \epsilon_{\bar{N}+1,\ell} > 0 \text{ or }\bar{N} = \ell, \text{ with $\ell$ odd}
     \end{cases} \, . 
\end{equation}

\ni We limit our analysis to $-2 < \mu < 2$: indeed $\mu \leq -2$ and $\mu \geq 2$ correspond to zero-filling and to full-filling (or quasi-full in case $\ell$ is odd), respectively.

The previous discussion implies that $E(\ell,\mu,J) = J(\ell-1) + \sum_{n=1}^{N(\ell,\mu)} \epsilon_{n,\ell}$ is a continuous, piecewise linear and increasing function of the chemical potential $\mu$:
\begin{align}
\label{app:eqn:enlm}
    E(\ell,\mu,J) &= J(\ell-1) - \mu N(\ell,\mu)  - 2 \, \sum_{n=1}^{N(\ell,\mu)} \cos(\frac{n \pi}{\ell+1}) = J(\ell-1) - \mu N(\ell,\mu) + 1- \csc(\frac{\pi}{2+2\ell})\sin(\frac{2 N(\ell,\mu) \pi + \pi}{2 + 2\ell}) \, .
\end{align}

\ni Indeed, we notice that in each interval of $\mu$ values for which $N(\ell,\mu)$ is constant, the contribution of the kinetic term is also constant and the energy grows linearly with the chemical potential $\mu$, with a slope given by $N(\ell,\mu)$. Continuity is recovered as, when $N(\ell,\mu)$ jumps from a value $N$ to $N+2$, which happens when $\epsilon_{N+1,\ell} + \epsilon_{N+2,\ell} = 0$, namely at $\mu = -\cos(\frac{(N+1) \pi}{\ell+1}) - \cos(\frac{(N+2)\pi}{\ell+1})$, the energy changes by $\Delta = - 2 \mu - 2 \left[ \cos(\frac{(N+1) \pi}{\ell+1}) + \cos(\frac{(N+2)\pi}{\ell+1}) \right] = 0$. 
The ground-state configuration at all values of $J$ and $\mu$ is then found by comparing the energies of all possible partitionings of the lattice:
\begin{equation}
\label{app:eqn:classical_gs}
    \mathcal{E}_{\rm GS}(\mu,J) = \min_{\{\ell_i\} \in \mathcal{P}(L)} \sum_i E(\ell_i,\mu,J) \, ,
\end{equation} 

\ni where $\mathcal{P}(L)$ denotes the set of all possible partitions of a chain of size $L$ and $\{\ell_i\}$ is one of these partitionings. The ground state configurations are reported in Fig.(4) of the main text, for $L=60$.

\begin{figure}
    \includegraphics[width = 0.5\linewidth]{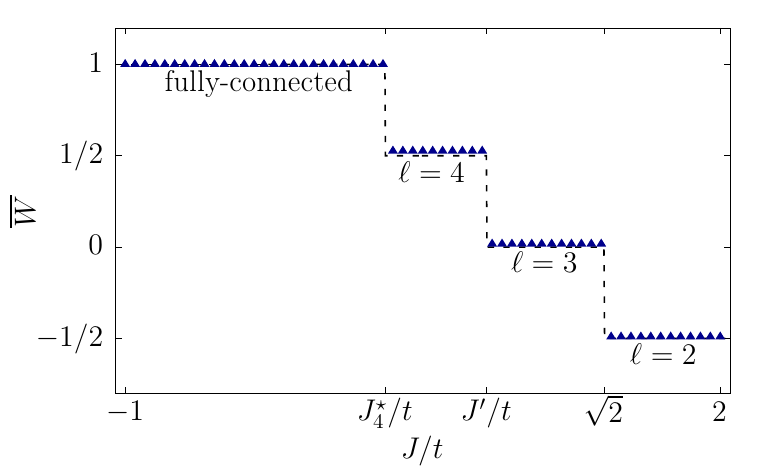}
    \caption{\textit{Chain-partitioning at $\nu = \frac{1}{2}$}. We plot $\overline{W} = \sum_i \langle W_{\bigcirc_{i_\ell}}\rangle/(L-1)$ as a figure of merit for the chain-partitioning at $h=0$, where the system size is $L=60$ and the filling fraction is fixed to $\nu=\frac{1}{2}$. Individual subchains are separated from one another by states of flux $\pi$: in order of increasing $J/t$, we go from a fully connected chain, to coverings with clusters of four ($\ell=4$), three ($\ell=3$) and two ($\ell =2$) sites, each hosting two particles. The dashed line shows the behaviour in the thermodynamic limit.}
    \label{app:fig:half_partitioning}
\end{figure}

At a fixed value of the filling fraction $\nu = \frac{N}{L}$, the lowest energy configurations strictly depend on the value of $\nu$, due to commensurability reasons. As noted in the main text, for $\nu=\frac{2}{3}$, the chain gets covered in trimers for all values of $J$ in $\left[\frac{3\sqrt{3}}{\pi} - \sqrt{2},\sqrt{2}\right]$. The two particles in each trimer are in the lowest energy state of a chain of $3$ sites, namely $\ket{n=1,n=2} = f^\dagger_{n=1} f^\dagger_{n=2} \ket{\rm vacuum}$, with $E_{\text{trimer}} = 2J - \sqrt{2}$. For $J<\frac{3\sqrt{3}}{\pi} - \sqrt{2}$, the system becomes fully connected, while for $J>\sqrt{2}$ particles dimerize into clusters of two sites, each of which carries an energy $E_{\text{dimer}} = J$. The boundary between the trimerized and dimerized regimes is inferred by the condition that $E_{\text{trimer}} = E_{\text{dimer}}$, since AB cages in the two cases only have different sizes, but are present in the same amount. Instead, we pass from trimers to a fully-connected chain when $E_{\text{fully-connected}} = \frac{L}{3} E_{\text{trimer}}$, which provides $J = J^\star = \frac{3\sqrt{3}}{\pi} - \sqrt{2}$ in the thermodynamic limit. The latter condition stems from the fact that $L/3$ trimers are needed to fully cover the chain. Notice that $E_{\text{fully-connected}} = J(L-1) -\sum_{n=1}^N 2 \cos(\frac{n \pi}{L+1}) \xrightarrow[L \to \infty]{} J(L-1) -\frac{L+1}{\pi} \int_0^{\nu \pi} dk \, 2 \cos(k) \approx L \left( J - \frac{2}{\pi} \sin(\nu \pi) \right)$.

We point out that, while an unconstrained particle density would lead to a covering with fourmers (clusters of four sites and two particles in each) below an intermediate value of $J$ in the aforementioned range -- see Fig.(4) of the main text, such covering is incommensurable with $\nu = \frac{2}{3}$. Nevertheless, in the half-filled regime $\nu=\frac{1}{2}$, increasing $J$ above zero leads from a fully connected chain to, in order, clusters of size $\ell = 4,3,2$ - see Fig.\ref{app:fig:half_partitioning}. In this case, the boundaries $J^\star_4$ and $J'$ between the regime of fourmers and the regimes in which the chain is either fully-connected or covered in trimers, are found by the conditions that $E_{\text{fourmer}} = E_{\text{trimer}}$ and $\frac{L}{4} E_{\text{fourmer}} = E_{\text{fully-connected}}$, respectively. We find: $J' = \sqrt{5}-\sqrt{2}$ and $J^\star_4 = \frac{8}{\pi}-\sqrt{5}$.

To find the ground-state configuration at $h=0$ numerically, we considered the following approaches. For a chain of size $L$ and an unconstrained filling fraction, we take into account the set of all inequivalent integer partitionings of the integer $L$, $\mathcal{P}(L)$, where by inequivalent it is meant that they do not differ by a mere reordering of the individual partitions. Each element of this set corresponds to the lengths $\{\ell_i\}$ of the AB cages in which the chain is fragmented. For each value of the chemical potential $\mu$ (in units of $t$, here set to $1$), the particle filling of each partition is then fixed by the relation \eqref{app:eqn:maxinteger} and the total energy for the partitioning $\{\ell_i\}$ is found as $\sum_i E(\ell_i,\mu,J)$ -- see Eq.\eqref{app:eqn:enlm}. Comparison of the energies associated to all the elements of $\mathcal{P}(L)$ yields the ground-state configuration.

In the cases in which the filling fraction is fixed to a value $\nu = N/L$, we instead consider all inequivalent partitions of the integer $N$ with only even integers $\{N_i\} \in \mathcal{P}_{\rm even}(N)$, corresponding to the occupation numbers of all the cages along the chain that contain particles. The sizes of such cages $\{\ell_i\}$ that minimize the total energy is then found \textit{for every element} $\{N_i\}$ of $\mathcal{P}_{\rm even}(N)$. The ground-state configuration will hence be specified by the particles' distribution and by the lengths of the cages containing them, which correspond to the lowest possible energy. Let the generic element of $\mathcal{P}_{\rm even}(N)$ be written as $\{N_1,N_2,\dots,N_M\} | \sum_{m=1}^M N_m = N$, where $M$ is the number of even integers that partition $N$. The steps are summarized as follows:
\begin{enumerate}
    \item The first $N_1$ particles are taken into account, which are contained in an AB cage whose size is allowed to vary in the range $\ell_1 \in [N_1,L-\sum_{m=2}^M N_m] = \mathcal{L}_1$. The limiting values of the latter interval correspond to the minimum and maximum available space: $\ell_1 = N_1$ coincides with the particles entirely filling the AB cage to which they belong, with no empty sites interspersed among them. The maximum size of $\ell_1$, on the other hand, is attained when all the other particles are squeezed in the remaining part of the chain, with no sites left unoccupied. For each value of $\ell_1 \in \mathcal{L}_1$, representing our partial configurations, we calculate the corresponding partial energies $\mathcal{E}_1 = \{E(\ell_1,N_1,J) | \ell_1 \in \mathcal{L}_1\}$.
    \item If $M>1$, we iteratively consider all the remaining $N_m$. Proceeding from $m=2$, $N_2$ particles will occupy an AB cage of $\ell_2$ sites. For each value of $\ell_1 \in \mathcal{L}_1$, $\ell_2$ will be allowed to vary in $[N_2,L-\ell_1-\sum_{m=3}^M N_m] = \overline{\mathcal{L}}_2(\ell_1)$. At this point, for every possible partial length $\ell_1 + \ell_2$ that takes values in a set which we denote as $\mathcal{L}_2$, we find the minimum energy, so that our partial configurations are now pairs $(\ell_1,\ell_2)$, each summing to a different partial length in $\mathcal{L}_2$ and that correspond to the partial energies $\mathcal{E}_2 = \left\{ \min\limits_{\ell_1+\ell_2=\bar{\ell}} \, [E_1(\ell_1,N_1,J) + E(\ell_2,N_2,J)] \Big| \bar{\ell} \in \mathcal{L}_2, \text{with } \ell_2 \in \overline{\mathcal{L}}_2(\ell_1) \, \text{and }  \ell_1 \in \mathcal{L}_1\right\}$. By finding the minimum of the energy for each value of the partial length $\ell_1+\ell_2$, we are ruling out many other configurations in which the cages sum to the same $\ell_1+\ell_2$, but that do not need to be taken into account, as they would remain at a higher energy when adding extra particles at the next iteration.
    \item At the next step, $N_3$ particles are added in an AB cage of size $\ell_3$. For every total partial length $\bar{\ell}$ in $\mathcal{L}_2$, $\ell_3$ will take values in $[N_3,L-\bar{\ell}-\sum_{m=4}^M N_m] = \overline{\mathcal{L}}_3(\bar{\ell})$. For every possible different value of $\bar{\ell} + \ell_3$, belonging to a set that we label as $\mathcal{L}_3$, with $\ell_3 \in \overline{\mathcal{L}}_3(\bar{\ell})$ and $\bar{\ell} \in \mathcal{L}_2$, we again find the minimum energy. The partial configurations are updated to tuples $(\ell_1,\ell_2,\ell_3)$ summing to the different elements of $\mathcal{L}_3$ and corresponding to the partial energies in a set $\mathcal{E}_3$.
    \item Proceeding in the same fashion, after the last iteration, we are left with partial configurations which are different M-tuples $(\ell_1,\dots,\ell_M)$ summing to distinct partial lengths in a set $\mathcal{L}_M$ and corresponding to the total energies in a set $\mathcal{E}_M$. Now, if $J \geq 0$, for all the empty sites which might have remained unoccupied after the described iterations, it is energetically favourable to remain disconnected from one another. Being separated by $\pi$-flux link states, they do not contribute to the energy, while, if they formed clusters, some of these $\pi$-fluxes would become $0$-flux states that give positive contribution to the total energy. Therefore, in this case, the minimum energy configuration corresponds to the M-tuple $(\ell_1,\dots,\ell_M)$ that attains the minimum value of $\mathcal{E}_M$. On the other hand, if $J<0$, possible empty sites will prefer to form the biggest available cluster to lower the energy. The partial energies $\mathcal{E}_M$ then need to be modified to $\mathcal{E}'_M$ = $\mathcal{E}_M + \{J(L-\bar{\ell}-1) | \bar{\ell} \in \mathcal{L}_M\}$, $L-\bar{\ell}$ being the size of the largest available empty cluster for each $\bar{\ell} \in \mathcal{L}_M$. The lowest energy configuration will then be obtained by finding $\min [\mathcal{E}'_M]$.
    \item Repeating the same procedure for each element of $\mathcal{P}_{\rm even}(N)$, each particle configuration will be associated to the best arrangement of cages, which in turn corresponds to the lowest possible energy for that specific distribution of the particles. The ground-state will then be given by the configuration that, among them, has minimum energy.
\end{enumerate}

\section{Effective dynamics for small $h/t,h/J$}
\label{app:sec:small_h}
As discussed in the main text, quantum fluctuations induced by $h\neq0$ restore tunneling between the disconnected partitions of the chain and provide the AB cages with a non-trivial dynamics. The effects of a small electric field are here addressed by deriving the effective dynamics stemming from a second-order perturbative expansion in orders of $h/t$, $h/J$. We will assume to work inside a range of parameters in which the system forms cages of size $\ell$, each containing two particles, which will occupy the lowest single particle energy levels, i.e. those corresponding to $n=1,2$ -- see the previous section. Here the number of particles is assumed to be fixed, so that we can forget about the chemical potential $\mu$. Given that a subchain of length $\ell$ containing $N$ particles has an energy $E_{\ell,\{n_1,\dots,n_N\}} = J(\ell-1) - \sum_{n \in \{n_1,\dots,n_N\}} 2 \cos(\frac{n \pi}{\ell+1})$, each two-particle cage will be associated to 
the energy $E_\ell = E_{\ell,\{1,2\}}$. Particles inside each cage are described by the wave-function $\bra{j_1,j_2}\ket{n_1 = 1,n_2 = 2}_\ell = \bra{j_1,j_2} f^\dagger_{n_1=1} f^\dagger_{n_2 = 2} \ket{0}_\ell$. In the limit in which only one kind of partition of size $\ell$ is stable, in the sense that all other configurations lie at considerably higher energies with respect to the matrix elements of the perturbation $V = h \sum_i S^x_{i_\ell}$, the effect of the electric field is seen in second order processes. Indeed, by turning $0$-flux states into $\pi$-flux states according to $S^x |\Phi^+\rangle = |\Psi^+\rangle$ and viceversa, the electric field can expand or contract a cage, as well as breaking or merging them, thus it cannot contribute at first order in perturbation theory, as its action goes out of the subspace in which all clusters have size $\ell$. Denoting the states belonging to the degenerate manifold of $\ell$-sites clusters as $\ket{\ell,g}$, where $g$ compactly indexes all the possible degenerate configurations due to the different distribution of clusters along the chain, the effective Hamiltonian determining the perturbative dynamics has matrix elements $(\mathcal{H}_\ell)_{g,g'} = \sum_{\ket{m} \notin \{\ket{\ell,g}\}} \frac{\bra{\ell,g'} V \ket{m} \bra{m} V \ket{\ell,g}}{E_\ell - E_m}$. Here $\{\ket{m}\}$ denotes a set of intermediate states which are connected by the perturbation $V$ to the ground state manifold $\{\ket{\ell,g}\}$. Depending on the nature of these intermediate states, we will have different possible second-order processes. Denoting as $M_\ell$ the total number of $\ell$-sites partitions and as $v_\ell$ the number of pairs thereof which are nearest-neighbours, all the possible outcomes are listed in the following:
\begin{enumerate}
    \item Links in the $\pi$-flux state $|\Psi^+\rangle$, which are neither neighboring nor contained in any cluster are flipped back and forth, resulting in identical processes that return back to the initial configuration. The intermediate states $\ket{m}$ differ from $\ket{\ell,g}$ only by a localized flip and have energy $E_m = E_\ell + J$. The contribution to the effective Hamiltonian then amounts to 
    \begin{equation}
    \label{app:eqn:contribution_1}
        -\frac{h^2}{J} [(L-1) - (\ell-1)M_\ell - (2 M_\ell - v_\ell)]=-\frac{h^2}{J} (L-1 - (\ell+1)M_\ell + v_\ell).
    \end{equation}
    
    \ni The term in parenthesis counts the number of such link states, for a chain terminating with sites: $(L-1)$ is the total number of links, $(\ell-1)M_\ell$ counts only those contained inside the clusters, while $(2M_\ell - v_\ell)$ those that are neighbouring individual clusters or pairs thereof.
    \item Links in the $|\Psi^+\rangle$ state which are neighbouring \textit{only one cluster} are flipped back and forth. The partition is then extended to include $\ell+1$ sites, while the particles will explore the spectrum of the enlarged subchain, given by the energies $E_{\ell+1,\{n_1,n_2\}}$ for varying quantum numbers $n_1,n_2 \in \{1,\dots,\ell+1\}$. At second order, we go back to the initial configuration. The matrix elements of the perturbation reduce to either
    \begin{equation}
    \label{app:eqn:enlarge_r}
        A_R(n_1,n_2) = {}_{\ell+1}\bra{n_1,n_2}\ket{(1,2)_\ell \, , \bullet} = \sum_{\substack{j_1,j_2 = 1 \\ j_1 < j_2}}^\ell {}_{\ell+1} \bra{n_1,n_2}\ket{j_1,j_2} \bra{j_1,j_2}\ket{1,2}_\ell
    \end{equation}

    \ni or
    \begin{equation}
    \label{app:eqn:enlarge_l}
        A_L(n_1,n_2) = {}_{\ell+1}\bra{n_1,n_2}\ket{\bullet \, ,(1,2)_\ell} = \sum_{\substack{j_1,j_2 = 2 \\ j_1 < j_2}}^{\ell+1} {}_{\ell+1} \bra{n_1,n_2}\ket{j_1,j_2} \bra{j_1-1,j_2-1}\ket{1,2}_\ell
    \end{equation}
    
    \ni depending on whether the electric field perturbation acts on the right or on the left of the initial cage. We have indicated $\ket{n_1,n_2}_{\ell+1}$ as the state of the enlarged cluster, while $\ket{(1,2)_\ell \, , \bullet}$ ($\ket{\bullet \, ,(1,2)_\ell}$)is the state of the original $\ell$-size cage, neighboured on the right (left) by an empty disconnected site, denoted with the $\bullet$ symbol. By symmetry, $\abs{A_L(n_1,n_2)} = \abs{A_R(n_1,n_2)} = \abs{A(n_1,n_2)}$. The total contribution to the energy then amounts to 
    \begin{equation}
        \label{app:eqn:contribution_2}
        2 \, h^2 (M_\ell - v_\ell) \sum_{\substack{n_1,n2 = 1\\ n_1 < n_2}}^{\ell+1} \frac{\abs{A(n_1,n_2)}^2}{E_{\ell,2} - E_{\ell+1,\{n_1,n_2\}}} \, ,
    \end{equation}
    
    \ni where the term $2 (M_\ell - v_\ell)$ counts the number of links that neighbour single clusters.
    \item A $\pi$-flux link neighbouring a single cluster is again flipped, resulting in a larger partition of $(\ell+1)$ sites. While in the previous case one returns to the initial configuration by flipping back the same link state, which lies either at the left or at the right end of the enlarged partition, now $V$ is made to act on the link at the opposite end. As a result, the $\ell$-sites cluster hops either backwards or forward by one site, with amplitude
    \begin{equation}
        \label{app:eqn:contribution_3}
        h^2 \sum_{\substack{n_1,n2 = 1\\ n_1 < n_2}}^{\ell+1} \frac{A_L(n_1,n_2) A_R(n_1,n_2)}{E_{\ell,2} - E_{\ell+1,\{n_1,n_2\}}} \, ,
    \end{equation}

    \ni where we have used equations \eqref{app:eqn:enlarge_r} and \eqref{app:eqn:enlarge_l}.
    \item $V$ acts between nearest-neighbouring partitions, that are merged into a single unit with $2\ell$ sites and four particles and then disconnected, at second order, into the same initial configuration. The matrix elements of the perturbation amount to: 
    \begin{align}
    \label{app:eqn:nn_clusters}
        \nonumber
        B(n_1,n_2,n_3,n_4) &= {}_{2\ell} \bra{n_1,n_2,n_3,n_4} \ket{(1,2)_\ell,(1,2)_\ell} \\
        & = \sum_{\substack{j_1,j_2 = 1 \\ j_1 < j_2}}^\ell \, \sum_{\substack{j_3,j_4 = \ell+1 \\ j_3 < j_4}}^{2\ell} {}_{2 \ell} \bra{n_1,n_2,n_3,n_4}\ket{j_1,j_2,j_3,j_4} {}_\ell \bra{j_1,j_2}\ket{1,2}_\ell \bra{j_3-\ell,j_4-\ell}\ket{1,2}_\ell \, ,
    \end{align}    
    
    \ni where $\ket{n_1,n_2,n_3,n_4}_{2\ell}$ is the four particles state in the enlarged cluster. The total contribution to the energy is then given by 
    \begin{equation}
        \label{app:eqn:contribution_4}
        h^2 v_\ell \sum_{\substack{n_1,\dots,n_4 = 1 \\ n_1<\dots<n_4}}^{2\ell} \frac{\abs{B(n_1,n_2,n_3,n_4)}^2}{2 E_{\ell,2} - E_{2\ell,\{n_1,n_2,n_3,n_4\}}} \, .
    \end{equation}
    \item $V$ acts on links enclosed by a cage, breaking it into two units, one of which remains empty of particles, by gauge-invariance. At second order, if the flipped link was in the bulk of the cluster, the only possibility is that $V$ acts back at the same position, leading to an overall constant contribution to the energy which is proportional to $M_\ell$ and is thus unimportant at fixed $M_\ell$. Differently, if the flipped link was at one of the edges, the initial partition is shortened to $\ell-1$ sites and, as $V$ acts again on the opposite edging link, the cluster hops either forward or backwards. The matrix elements of $V$ that enter in these processes are given by
    \begin{equation}
        C_R(n_1,n_2) = \bra{(n_1,n_2)_{\ell-1} \, , \bullet}\ket{1,2}_\ell = \sum_{\substack{j_1,j_2 = 1 \\ j_1<j_2}}^{\ell-1} {}_{\ell-1} \bra{n_1,n_2}\ket{j_1,j_2} \bra{j_1,j_2}\ket{1,2}_\ell \, ,
    \end{equation}

    \ni and 
    \begin{equation}
        C_L(n_1,n_2) = \bra{\bullet \, ,(n_1,n_2)_{\ell-1}}\ket{1,2}_\ell = \sum_{\substack{j_1,j_2 = 1 \\ j_1<j_2}}^{\ell-1} {}_{\ell-1} \bra{n_1,n_2}\ket{j_1,j_2} \bra{j_1+1,j_2+1}\ket{1,2}_\ell \, ,
    \end{equation}

     \ni depending on whether $V$ acts at the right or left edge of the initial cage. The contribution to the hopping amplitude hence amounts to
     \begin{equation}
     \label{app:eqn:contribution_5}
         h^2 \sum_{\substack{n_1,n_2 = 1 \\ n_1 < n_2}}^{\ell-1} \frac{C_L(n_1,n_2) C_R(n_1,n_2)}{E_{\ell,2}-E_{\ell-1,\{n_1,n_2\}}} \, .
     \end{equation}
     
     \ni We point out that, if $\ell = 2$, such processes do not contribute, as the only link contained in each two-site cluster is in the $\ket{\Phi^-}$ state and is thus annihilated by $V$.
\end{enumerate}
\begin{figure}[h!]
    \centering
    \includegraphics[width=0.5\linewidth]{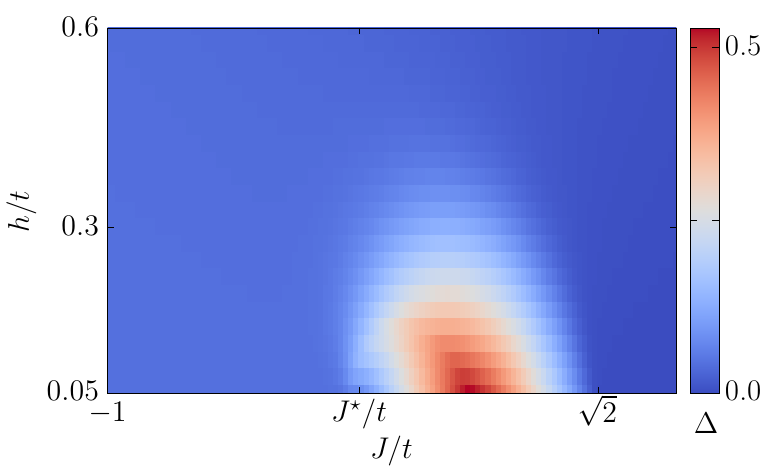}
    \caption{\textit{Spectral gap at $\nu = \frac{2}{3}$}. A spectral gap opens up in the Mott-insulating regime in which the lattice is completely covered by trimers. The size of the lattice is set to $L=120$.}
    \label{app:fig:2div3_spectral}
\end{figure}
Collecting all these processes together, the effective Hamiltonian that describes the second order perturbative dynamics of clusters of $\ell$ sites is given by (neglecting an additive constant):
\begin{equation}
\label{app:eqn:effham}
    \mathcal{H}_{\ell} = \mathcal{P}_\ell \sum_i \left\{ - t_{\text{eff},\ell} (b^\dagger_i b_{i+1} + H.c.) + V_{\text{eff},\ell} \, n_i^b n_{i+\ell}^b \right\} \mathcal{P}_\ell \, .
\end{equation}

\ni The bosonic operators $b_i$ annihilate a whole $\ell$-sized cluster originating from site $i$: two particles are destroyed and all the links enclosed by the cluster are set to $|\Psi^+\rangle$ states. The projectors $\mathcal{P}_\ell$ impose the hardcore constraints $b^\dagger_i b^\dagger_{i+n} = 0$ $\forall n \in \{0,\ell-1\}$, to account for the spatial extent of these cages. We furthermore notice that the bosonic filling is related to the density of particles $\nu$ by $\nu_m(\ell) = \frac{\nu}{2}$, as these composite bosons are mesons with two $\mathbb{Z}_2$-charges locked inside. The Hamiltonian parameters are obtained from the previous analysis as:
\begin{equation}
\label{app:eq:effpar}
\begin{aligned}
    t_{\text{eff},\ell} &= - h^2 \sum_{\substack{n_1,n2 = 1\\ n_1 < n_2}}^{\ell+1} \frac{A_L(n_1,n_2) A_R(n_1,n_2)}{E_{\ell,2} - E_{\ell+1,\{n_1,n_2\}}} - (1-\delta_{\ell,2}) \, h^2 \sum_{\substack{n_1,n_2 = 1 \\ n_1 < n_2}}^{\ell-1} \frac{C_L(n_1,n_2) C_R(n_1,n_2)}{E_{\ell,2}-E_{\ell-1,\{n_1,n_2\}}} \, ,\\ 
    V_{\text{eff},\ell} & = - \frac{h^2}{J} - 2 \, h^2 \sum_{\substack{n_1,n2 = 1\\ n_1 < n_2}}^{\ell+1} \frac{\abs{A(n_1,n_2)}^2}{E_{\ell,2} - E_{\ell+1,\{n_1,n_2\}}} + h^2 \sum_{\substack{n_1,\dots,n_4 = 1 \\ n_1<\dots<n_4}}^{2\ell} \frac{\abs{B(n_1,n_2,n_3,n_4)}^2}{2 E_{\ell,2} - E_{2\ell,\{n_1,n_2,n_3,n_4\}}} \, .
\end{aligned}
\end{equation}

As discussed in the main text, at the bosonic density $\bar{\nu}_m(\ell) = \frac{1}{1+\ell}$, which corresponds to the meson "half-filling" due to the constraints imposed by $\mathcal{P}_\ell$, the point $\delta_c$ at which $\delta_\ell = \frac{V_{\text{eff},\ell}}{2 \, t_{\text{eff},\ell}} = 1$ is a critical point for the model \eqref{app:eqn:effham}, since it signals a transition from a Luttinger Liquid ($\delta_\ell \leq 1$) to a Mott insulator ($\delta_\ell > 1$) \cite{alcaraz1999exactly}. At the filling $\nu = \frac{2}{3}$, for $J>\sqrt{2}$ and $h=0$ the system forms dimers, whose density corresponds to the "half-filling" value $\bar{\nu}_m(\ell = 2) = \frac{1}{3}$ for the composite bosons. At small $h$, dimers move according to Eq.\eqref{app:eqn:effham}, which is valid for $J$ sufficiently larger than $\sqrt{2}$, at which individual trimers and dimers have the same energy. Let us explicitly restore the units of $t$. Rigorously, it should be imposed that $\displaystyle{\frac{h}{J-\sqrt{2}t} \ll 1}$. Explicit evaluation of the parameters of the effective model provides:
\begin{align}
\label{app:eqn:l2hop}
    & t_{\text{eff},2} = \frac{h^2}{J} \frac{t^2}{J^2 - 2t^2} \, , \\
\label{app:eqn:l2int}
    &V_{\text{eff},2} = 2 \, t_{\text{eff},2} \, ,
\end{align}

\ni so that $\delta_2 = 1$. We can then conclude that the dimerized case lies exactly at the critical point between a Luttinger Liquid and a $\mathbb{Z}_3$ Mott-insulating behaviour. For $J \sim \sqrt{2}t$, the description based on Eq.\eqref{app:eqn:effham} fails and quantum fluctuations of the size of the cages due to $h \neq 0$ induce a transition to a Mott insulating behaviour in which trimers completely cover the chain. Such phase is characterized by the opening of both a charge gap (as reported in the main text) and a spectral gap -- see Fig.\ref{app:fig:2div3_spectral}.

\begin{figure}[h!]
    \centering
    \includegraphics[width=0.5\linewidth]{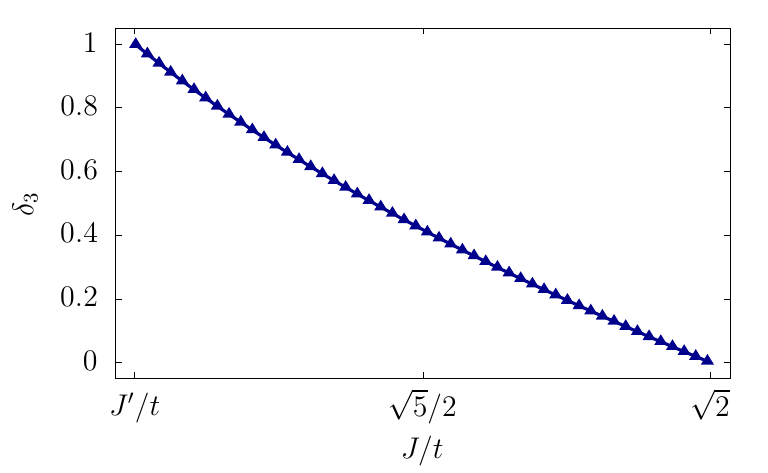}
    \caption{\textit{Ratio $\delta_3 = \frac{V_{\text{eff},3}}{2 t_{\text{eff},3}}$}. In this plot, the ratio between the effective parameters of the perturbative model \eqref{app:eqn:effham} in the trimerized regime is shown in the range $J \in (J',\sqrt{2}t)$. The perturbative expansion is valid sufficiently far away from the limiting values of the interval.}
    \label{app:fig:delta3}
\end{figure}

\begin{figure}[h!]
    \centering
    \includegraphics[width=0.9\linewidth]{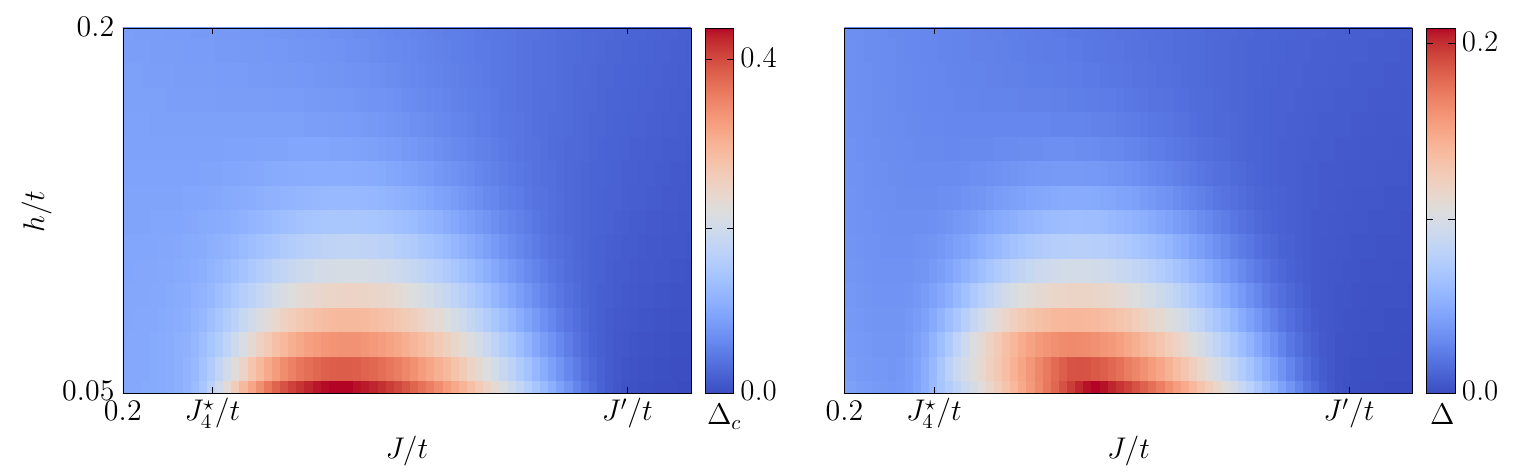}
    \caption{\textit{Charge and spectral gaps at $\nu = \frac{1}{2}$}. Both a charge gap (left panel) and a spectral gap (right panel) open up in the Mott-insulating regime in which the lattice is completely covered by fourmers. The size of the lattice is $L=120$.}
    \label{app:fig:1div2_gaps}
\end{figure}

For $\nu = \frac{1}{2}$, an equivalent analysis applies. The behaviour of the system at $h=0$ is reported in Fig.\ref{app:fig:half_partitioning}, where we distinguish regimes in which the chain remains fully-connected, or breaks into fourmers, trimers or dimers. The effective parameters for the $\ell=2$ cages will still be given by the equations \eqref{app:eqn:l2hop}, \eqref{app:eqn:l2int}. In the case of trimers, the effective bosonic density is set at its "half-filling" value $\bar{\nu}_m(\ell = 3) = \frac{1}{4}$. The ratio between the effective parameters of the dynamics $\delta_3$ is reported in Fig.\ref{app:fig:delta3}, and is valid in the range $J/t \in (\sqrt{5}-\sqrt{2},\sqrt{2})$, sufficiently far away from the edges of the interval, at which a single trimer becomes degenerate with a fourmer or a dimer. Namely we require $\displaystyle{\frac{h}{J-J'}}$,$\displaystyle{\frac{h}{J-\sqrt{2}t} \ll 1}$. We point out that $\delta_3 < 1$, so that trimers are strictly in the Luttinger Liquid regime of model \eqref{app:eqn:effham}. For $J \sim J' = (\sqrt{5}-\sqrt{2})t$, as in the previous case, the size of the cages fluctuates between trimers and fourmers configurations, determining a transition to a $\mathbb{Z}_4$ Mott insulating phase in which four-sites clusters cover the chain. Also in this instance, both a charge gap and a spectral gap are open -- see Fig.\ref{app:fig:1div2_gaps}.

\section{Analysis at small $\displaystyle{\frac{t}{h}}$}
\label{app:sec:large_h}
In this section, we address the large $\frac{h}{t}$ limit of the model \eqref{app:eqn:spin1red}. In the main text we have discussed the main features of this limit, understood as the regime at which pairs of particles are strongly confined due to the linearly growing attractive interaction caused by electric field strings connecting them. We shall demonstrate that this effect is valid for any $\frac{J}{t}$, which we allow to be comparable with $\frac{h}{t}$. We consider the fermionic rewriting of the model, obtained via a JW transformation that leads to Eq.\eqref{app:eqn:fermionic_spin1red}. The Hamiltonian can thus be split as a sum of the local, pure-gauge terms
\begin{equation}
    \mathcal{H}_0 = h \sum_i S^x_{i_\ell} + J \sum_i ( S^z_{i_\ell} )^2
\end{equation}

\ni plus the gauge-invariant matter kinetic term, which we treat as a perturbation
\begin{equation}
    V = t \sum_i (c^\dagger_i S^z_{i_\ell} c_{i+1} + H.c.) \, .
\end{equation}

\ni Here we want to work at fixed density $\nu$, hence we neglect the chemical potential contribution. Local diagonalization of $\mathcal{H}_{0,i}$ provides the three eigenvalues $\epsilon_\Uparrow = \frac{J}{2} + \sqrt{\frac{J^2}{4} + h^2}$, $\epsilon_0 = J$ and $\epsilon_\Downarrow = \frac{J}{2} - \sqrt{\frac{J^2}{4} + h^2}$, corresponding to the states $|\epsilon_{\Uparrow}\rangle = |\Psi^+\rangle + ({J+\sqrt{J^2+4h^2}}) |\Phi^+\rangle/{2h}$, $|\epsilon_0\rangle=|\Phi^-\rangle$ and $|\epsilon_{\Downarrow}\rangle=|\Psi^+\rangle+(J-\sqrt{J^2+4h^2}) |\Phi^+\rangle/{2h}$, respectively. The groundstate manifold of $\mathcal{H}_0$ will then be characterized by particles being tightly-packed into dimers and separated, according to gauge-invariance, by links in the $\ket{\Phi^-}$ states, which carry one unit of electric field as $E_{i_\ell} |\Phi^-_{i_\ell}\rangle = + |\Phi^-_{i_\ell}\rangle$. All the other bonds of the lattice will be in the lowest-energy $|\epsilon_\Downarrow\rangle$ states. The local effective Hilbert space of the link states will thus be two-dimensional, in analogy to the standard $\mathbb{Z}_2$ chain. The ground-state is highly-degenerate, due to all the possible distinct ways of arranging dimers along the chain. 

The perturbative action of $V$, which restores the dynamics of dimers, can be taken into account via a Schrieffer-Wolff transformation \cite{schrieffer1966relation,bravyi2011schrieffer}. We notice that $V$ has no block-diagonal action in the ground-state manifold, as it takes states out of the degenerate subspace. Therefore, the dynamics of dimers is recovered in a second order expansion in orders of $t/h$: second order processes have the effect of contracting and stretching the electric field lines connecting pairs of particles, determining the motion of dimers and their interactions, analogously to what has been demonstrated for the standard $\mathbb{Z}_2$ chain \cite{borla2020confined,kebric2021confinement}. To this order, the effective Hamiltonian reads
\begin{equation}
    H_{\text{eff}}^{(2)} = \frac{1}{2} \mathcal{P}_{\text{dimers}} [S^{(1)},V] \mathcal{P}_{\text{dimers}}
\end{equation}

\ni where $\mathcal{P}_{\text{dimers}}$ is a projector into the groundstate manifold and $S^{(1)}$ is anti-hermitian and is found from the condition that $[H_0,S^{(1)}] =V$:
\begin{equation}
    S^{(1)} = \text{i} \sum_i (c_i^\dagger c_{i+1} + H.c.) \left( \frac{t}{h} S^y_{i_\ell} + \frac{tJ}{h^2} (S^z_{i_\ell} S^y_{i_\ell} + S^y_{i_\ell} S^z_{i_\ell}) \right) \, .
\end{equation}

\ni We find that the effective dynamics is described by the same model as in the opposite limit of small $h/t,h/J$ -- see Eq.\eqref{app:eqn:effham} for $\ell = 2$, which then reads
\begin{equation}
\label{app:eqn:effham_large_h}
    H^{(2)}_{\text{eff}} = \mathcal{P} \sum_i \{- \tilde{t}_{\text{eff}} (b^\dagger_i b_{i+1} + H.c.) + \tilde{V}_{\text{eff}} \,  n_i^b n_{i+1}^b \} \mathcal{P}
\end{equation}

\ni with
\begin{equation}
    \tilde{t}_{\text{eff}} = -\frac{t^2}{2h^2} \frac{J^3 + 4Jh^2 + (J^2 - 2Jh - 2h^2)\sqrt{4h^2+J^2} + 4h^3 }{J^2-2h\sqrt{4h^2+J^2} + 4h^2}
\end{equation}

\ni and
\begin{equation}
    \tilde{V}_{\text{eff}} = 2 \tilde{t}_{\text{eff}} \, .
\end{equation}

\ni Following the discussion of the main text and of section \eqref{app:sec:small_h} of this Supplemental Material, we can conclude that, since $\tilde{\delta} = \frac{\tilde{V}_{\text{eff}}}{2\tilde{t}_{\text{eff}}} = 1$, the effective dynamics provided by Eq.\eqref{app:eqn:effham_large_h} lies at the critical point $\delta_c$ between a LL and a $\mathbb{Z}_3$ Mott insulator of dimers.

\end{document}